\documentclass[fleqn,usenatbib]{mnras}

\usepackage{newtxtext,newtxmath}

\usepackage[T1]{fontenc}

\DeclareRobustCommand{\VAN}[3]{#2}
\let\VANthebibliography\thebibliography
\def\thebibliography{\DeclareRobustCommand{\VAN}[3]{##3}\VANthebibliography}

\usepackage{graphicx}	
\usepackage{amsmath}	
\usepackage[dvipsnames]{xcolor}
\usepackage{subfig}
\usepackage{pifont}
\usepackage{booktabs}

\newcommand{\doublecheck}{\checkmark \kern-0.3em \checkmark}
\newcommand{\xmark}{$\times$}

\title[SE3D tests on mock observations]{SE3D: Testing the recovery of stellar population, dust and structural properties on mock-observed toy model and simulated galaxies}

\author[Junkai Zhang et al.]{
Junkai Zhang,$^{1}$
Steven Ramnichal,$^{2}$
Stijn Wuyts,$^{2}$\thanks{E-mail: s.wuyts@bath.ac.uk}
Cheng Li$^{1}$
\\
$^{1}$Department of Astronomy, Tsinghua University, Beijing 100084, China\\
$^{2}$Department of Physics, University of Bath, Claverton Down, Bath, BA2 7AY, UK\\
}

\date{Accepted 2026 May 2. Received 2026 May 1; in original form 2025 November 24}

\pubyear{2026}

\begin{document}
\label{firstpage}
\pagerange{\pageref{firstpage}--\pageref{lastpage}}
\maketitle

\begin{abstract}
The translation from direct observables to physical properties of galaxies is a key step in reconstructing their evolutionary histories. Star-dust geometry and inhomogeneous stellar populations can induce spatial variations in the mass-to-light ratio, complicating this process. In this paper, we present tests of {\tt SE3D}, a novel modelling framework built around a radiative transfer emulator, aimed at tackling this problem.  We test the ability to recover known intrinsic properties of toy model and TNG50 simulated galaxies by jointly fitting mock observations of their multi-wavelength photometric and structural properties.  We find an encouraging performance ($\lesssim$ 0.1 dex) for several key characteristics, including the bulk stellar mass, dust mass and SFR, as well as their respective radial extents.  We point out limitations, and investigate the impact of various sources of model mismatch.  Among them, mismatch in the shapes of star formation histories contributes most, with radial and azimuthal structure and stellar metallicity distributions playing a progressively more minor role.  We also analyse the evolution from $z=2$ to $z=0$ of resolved stellar and dust properties of TNG galaxies, as measured intrinsically and expressed in their distribution across $UVJ$ and IRX-$\beta$ diagnostic diagrams.  We test different methods to assign dust to the simulation, and find a persistent lack of $M_{\rm dust}/M_{\rm star}$ evolution and a more limited dynamic range across the diagnostic diagrams compared to observations.

\end{abstract}

\begin{keywords}
galaxies: evolution -- galaxies: structure -- galaxies: stellar content -- galaxies: ISM -- ISM: dust, extinction -- galaxies: photometry
\end{keywords}



\section{Introduction}
\label{sec:introduction}

Since the late 1990s, deep lookback surveys have come a long way.  Initially, studies of distant galaxies were carried out by two largely disjoint communities focussing on either the rest-frame ultraviolet (UV) emission of relatively unobscured galaxies \citep[e.g.,][]{Williams1996, Madau1996, Steidel1996}, or the sub-mm emission of luminous and dusty `monsters' \citep[e.g.,][]{Hughes1998, Blain1999}.  The {\it Spitzer} and {\it Herschel} space telescopes played a critical role by filling the wavelength gap between these two extrema \citep{Fazio2004,Rieke2004,Poglitsch2010,Griffin2010}.  As such, they enabled a more complete and holistic census of the distant galaxy population, from unobscured to highly dust-enshrouded.  The range in wavelengths over which resolved diagnostics of distant galaxies are accessible has steadily increased too: first with the installation of the WFC3 camera onboard the {\it Hubble Space Telescope} (HST, e.g., \citealt{Conselice2011,Szomoru2011,Law2012}), subsequently thanks to (sub)mm interferometers such as ALMA and NOEMA \citep[e.g.,][]{Hodge2016,Hodge2019,Tadaki2017a,Tadaki2020,Tan2024,Pastras2025}, and ultimately with the launch of {\it JWST} \citep[][see, e.g., \citealt{Suess2022,Kartaltepe2023,HuertasCompany2024,Martorano2025} among many others]{Gardner2006}.

For galaxies at cosmic noon ($1 \lesssim z \lesssim 3$), these technological advances have increasingly enabled panchromatic studies, capturing flux and colour information as well as global structural parameters (albeit at a mix of resolutions) all the way across the rest-UV/optical/near-, mid- and far-infrared.  The observational landscape for wide-area multi-wavelength lookback surveys stands to undergo a transformation of its own, with the {\it Euclid} mission underway \citep{Aussel2025,Quilley2025}, Rubin-LSST receiving first light, and the {\it Roman} and {\it Chinese Space Station Telescope (CSST)} missions projected for launch in the next few years. Together, these will deliver galaxy samples of unprecedented size, with SEDs and basic structural measurements sampled out to the near-IR.

In an effort to exploit these increasingly rich observational datasets, we introduced in a companion paper the {\tt SE3D} framework \citep[][hereafter R26]{Ramnichal2026}.  {\tt SE3D} aims to ingest both Spectral Energy Distribution (SED) and structural information of galaxies, to infer their spatial distribution of stellar populations and dust.  As such, the method goes beyond conventional SED modelling \citep[see][for a review]{Conroy2013} and is more akin to approaches that adopt 3D dust radiative transfer (RT) to self-consistently model the emerging emission from a galaxy composed of a distribution of sources embedded in a dusty medium \citep[e.g.,][]{Popescu2000,Popescu2011,Efstathiou2000,Efstathiou2022,Williams2019,Verstocken2020,Nersesian2020}.  One key distinction with respect to the latter is the use of a machine learning (ML) emulator.  Once trained on a set of actual RT calculations, the emulator is capable of translating a set of input parameters\footnote{Input parameters include the stellar and dust mass content, parameters describing the spatial distributions of these respective components including their size, thickness and radial density profile, the star formation history (log-normal at any radius) and gradient thereof, and viewing angle under which the galaxy is observed.  For a full description, we refer the reader to Table 1 in R26. \label{param.footnote}} defining the physical make-up of a galaxy to the corresponding output observables, i.e., wavelength-dependent fluxes, half-light radii and light profiles (see R26 for details).  In this context, we use the term Spectral Distribution (SD) to refer to the characterisation of the wavelength dependence of any such observable. If referring specifically to the wavelength dependence of a measure of galaxy structure (e.g., its effective radius or S\'{e}rsic index), we will use the term Spectral Structural Distribution. The emulator's computational efficiency then allows its use within a Bayesian fitting routine.  The data being fit to would nominally comprise both the galaxy-integrated SED (i.e., flux SD) as well as the available Spectral Structural Distributions (e.g., size and S\'{e}rsic index SDs).

In this paper, we verify {\tt SE3D}'s ability to recover known intrinsic quantities from mock observations with finite sampling in terms of wavebands and finite accuracy of photometric and structural measurements.  Beyond the application to mock-observed toy model galaxies, we also explore its use on mock observations generated from galaxies in the TNG50 cosmological simulation.  These simulated galaxies feature more complexity in terms of their star formation history (SFH), stellar and dust distribution.  The use of simulated galaxies to test the recovery of physical quantities via SED modelling was previously explored by, e.g., \citet{Wuyts2009a,Mitchell2013,Hayward2015,Lower2020,Gilda2021,Cochrane2025}.  Using our test results, we consider the physical conditions that make for a better/poorer recovery on such simulated galaxies.

As part of our analysis, we also quantify the distribution of age gradients, relative dust versus stellar radii, and thickness of the dust and stellar disks in massive star-forming galaxies (SFGs) in TNG50.\footnote{We likewise consider the impact of different choices involved in `painting' dust onto the simulation in post-processing, as this component of the interstellar medium is not explicitly modelled in TNG.}  We evaluate their distribution in key observational diagrams such as the rest-frame $UVJ$ and IRX-$\beta$ planes.  We contrast the spread of TNG galaxies across those diagnostic diagrams to the distribution of toy model galaxies on which the ML emulator was trained, and to the distribution of observed galaxies since cosmic noon.  In both $UVJ$ and IRX-$\beta$, we identify subsets of the observed galaxy population without counterparts among RT-processed TNG galaxies, but for which toy model galaxies with similar characteristics exist in our library.  We highlight how vectors of increasing $E(B-V)$ and increasing $A_V$ have different slopes in the $UVJ$ plane, and discuss the variety of physical conditions (including star-dust geometries) responsible for the considerable spread of observed galaxies in the IRX-$\beta$ plane.

Our paper is structured as follows.  Section\ \ref{sec:datasample} lays out the sample selection of TNG50 galaxies and the definition of toy models considered in this work.  Section\ \ref{sec:methodology} addresses how mock observations are produced for them, and recaps the {\tt SE3D} fitting methodology to which both toy model and simulated galaxies will be subject.  As not all parameters that are fit for have a hard ground truth in the simulated galaxy case, we also elaborate on how `truth' is defined for TNG galaxies.  In Section\ \ref{sec:results}, we present the results, consisting of intrinsic and observed properties of TNG50 galaxies, and the recovery exercise on toy model and simulated galaxies. We discuss the implications of our results in Section\ \ref{sec:discussion}.  Finally, we summarize our conclusions in Section\ \ref{sec:summary}.

Throughout this work, we adopt a \citet{Chabrier2003} stellar initial mass function (IMF) and a flat $\Lambda$CDM cosmology with $\Omega_{\Lambda} = 0.7$, $\Omega_m = 0.3$ and $H_0 = 70\ {\rm km}\ {\rm s}^{-1}\ {\rm Mpc}^{-1}$.

\section{Data}
\label{sec:datasample}

\subsection{IllustrisTNG}
\label{sec:TNG}

We make use of the highest-resolution run of the TNG suite of hydrodynamical cosmological simulations: TNG50 \citep{Nelson2019b}.  TNG50 has a mass resolution of $8.5 \times 10^4\ M_{\odot}$ ($4.5 \times 10^5\ M_{\odot}$) for baryonic (dark matter) particles.  Executed using the moving mesh code AREPO \citep{Springel2010}, TNG solves the gravitational and hydrodynamic forces, and pairs this with subgrid recipes for astrophysical processes such as star formation, accretion onto supermassive black holes, and the feedback arising from either of these.  Doing so, it has been shown to produce an evolving galaxy population which shares many of the characteristics seen among galaxies in the real Universe, from their distribution of masses to star formation rates and global morphological and/or dynamical properties \citep[see e.g.,][]{Pillepich2019, Zanasi2021, Nelson2021}.  Especially of use to our {\tt SE3D} tests, within the cosmological box of approximately 50 Mpc side length it features galaxies with a range of star formation histories, structural properties, stellar age gradients and relative distributions of stars and interstellar medium (ISM).  Moreover, since these galaxy properties arise self-consistently within the cosmological context of the simulation, all of them are more complex in nature than the simplified forms we adopt when setting up the toy model galaxies employed in our modelling. In other words, even if TNG50 galaxies are not fully representative of galaxies in the real Universe (see Section\ \ref{sec:obsdiagrams}), they can still serve as a useful benchmark to test how well basic properties can be recovered from complex systems when treated with simplified toy models.

\subsection{TNG50 sample selection}
\label{sec:sample}

We extracted samples of simulated massive SFGs from three snapshots of TNG50, corresponding to redshifts $z = 0$, 1 and 2.  The adopted stellar mass limit of $10^{10}\ M_{\odot}$ (within 30 pkpc from the galaxy centre) was imposed to ensure sufficient resolution as intrinsic shapes and bulge-to-total ratios for lower mass (and to some degree higher redshift) galaxies may be affected by resolution effects \citep{Zhang2022, Xu2024, Zeng2024}. The star-forming nature was evaluated by requiring a specific star formation rate ${\rm sSFR} \equiv \frac{\rm SFR}{M_*} > \frac{1}{3 t_{\rm H}}$, where $t_{\rm H}$ represents the Hubble time at the redshift of the galaxy. Our focus on SFGs over quiescent galaxies stems from our motivation to study the impact of dust, which is more prevalent in the ISM of SFGs.  We thus obtained samples of = 521, 566 and 289 SFGs at $z = 0$, 1 and 2, respectively.  The fact that the sample size does not rise monotonically with decreasing redshift simply reflects that, while new SFGs cross the stellar mass limit, others quench and drop below the sSFR criterion. We note that some galaxies, or more precisely progenitors/descendants, enter our sample at more than one redshift: 374 appear twice, and 52 appear in all three of the considered redshifts.  We do not consider this an issue, as the time interval between selected snapshots is large compared to the relevant timescales on which the galaxies evolve (e.g., their rotation period, depletion time, or inverse of their specific star formation rate).  For the purpose of our recovery analysis, descendants of galaxies already included at a higher redshift can thus safely be regarded as useful test objects with largely independent characteristics compared to their higher-redshift progenitor system.

\subsection{Toy model galaxies}
\label{sec:toymodels}

Other than simulated galaxies, we will consider in this work also populations of toy model galaxies with a range of stellar population and dust properties.  These form the basis of the SKIRT RT library on which the {\tt SE3D} ML emulator was trained.  The parametrization of the stellar population and dust components was introduced in detail by R26 (see also Footnote\ \ref{param.footnote}), with their Table 1 summarizing the parameter distributions used to generate the training library.  The observational characteristics of this library will be displayed, alongside TNG50 and actual observed galaxy distributions, in Section\ \ref{sec:prop}.

A good recovery of intrinsic properties from mock observations of such toy model galaxies is not guaranteed, for several reasons.  First, the mock observations are generated using actual SKIRT RT, whereas the {\tt SE3D} fitting (Section\ \ref{sec:framework}) is employing an emulator to predict the spectral distributions associated with a set of physical parameters.  While R26 demonstrated a satisfactory emulator performance, its ability to reproduce SKIRT ground truth is not perfect.  Secondly, observational limitations in the form of finite wavelength sampling and measurement uncertainties could leave open degeneracies between different parameters describing a toy model.

We will therefore also carry out a recovery analysis on a set of mock-observed toy model galaxies not previously seen by the emulator (Section\ \ref{sec:recov_toy}).  We generate this test set drawing from the same parameter distributions defined in R26's Table 1, with a few exceptions.  First, in analogy to our TNG50 analysis, we generate the mock observations for three discrete redshifts: $z = 0$, 1 and 2, fixing the age since the onset of star formation (Age) to the age of the Universe at that redshift.  Secondly, we create our test galaxies to have flattened disk geometries, with thickness of their stellar and dust disks ($C_{\rm star}/R_{\rm star}$ and $C_{\rm dust}/R_{\rm dust}$, respectively) drawn (independently) from a truncated Gaussian distribution with mean and standard deviation $(\mu,\ \sigma) = (0.2,\ 0.1)$ bounded by $[0.1,\ 1]$.

In total, we run SKIRT on 600 such toy models, equally split between $z = 0$, 1 and 2.  To each, we assign a random viewing angle on a sphere by drawing from a flat distribution in $\cos(\theta)$, corresponding to more edge-on than face-on inclinations.

\section{Methodology}
\label{sec:methodology}

\subsection{From TNG galaxies to mock observations}
\label{sec:TNG2mockobs}

\subsubsection{Assigning dust to TNG galaxies}
\label{sec:assigningdust}

Since dust physics is not explicitly incorporated within TNG, we employ an ad hoc approach to assign dust mass to a subset of the gas particles.  Two questions therefore need to be addressed: which gas particles have dust associated with them, and how much.  We followed \citet[][see also \citealt{Gebek2025}]{Torrey2012} in identifying dust-hosting gas particles as those that are sufficiently cold or dense:
\begin{equation}
    \log_{10} \left( \frac{T}{[{\rm K}]} \right) < 6+0.25 \log_{10} \left( \frac{\rho}{10^{10}[M_\odot h^2 {\rm kpc}^{-3}]} \right)
\end{equation}
To them, we assign dust by an amount based on the particle's gas mass and metallicity.  Specifically, we adopt the metallicity-dependent dust-to-gas mass ratio empirically calibrated by \citet{DeVis2019}, imposing a maximum $M_{\rm dust}/M_{\rm gas} = 0.01$ to avoid extrapolation.  As variations on these choices exist in the literature, we evaluate their impact in Appendix\ \ref{app:dustassignment}.

\subsubsection{Dust radiative transfer}
\label{sec:RT}

Following the procedure detailed in R26, we use SKIRT \citep{Camps2020} to calculate the effects of scattering, absorption, and emission in a dusty medium starting from the tabulated input properties of star and dust particles. The stellar particle table contains coordinates, masses, ages and metallicities of star particles belonging to a TNG galaxy, as well as the kernel size of each star particle, taken to be the smoothing length {\tt StellarHsml} of the respective particle in the TNG database.  The dust particle table contains coordinates of TNG gas particles, their associated dust mass, and kernel sizes (again taken as the smoothing length obtained from TNG particle-level data).  To facilitate the RT calculation within SKIRT, a 3D dust density field is reconstructed on a meshed grid from the information in this dust particle table.  We use a Cartesian spatial grid spanning a 3D space of $20R_{\rm{star}}\times 20R_{\rm{star}}\times 20C_{\rm{star}}$, where $R_{\rm star}$ refers to the radius of a circle encompassing half the stellar mass in the $xy$ plane and $C_{\rm star}$ the height of an ellipse encompassing half the stellar mass in the averaged $xz$ and $yz$ projections.  We adopt a symmetric power-law distribution of mesh points, ensuring the most accurate radiative transfer calculation in the densest part of the galaxy. The sizes of grid cells grow symmetrically when moving outwards, with a size ratio between innermost and outermost cells equal to 30.

The source emission of stellar particles was interpolated from a grid of  \cite{Bruzual2003} Simple Stellar Population (SSP) templates spanning a range in age and metallicity. For the dust, we make use of the THEMIS dust model (The Heterogeneous dust Evolution Model for Interstellar Solids, \citealt{Jones2017}), which specifies the dust chemical composition and grain size distribution. 

Following \citet{Trayford2017}, we resample stellar particles younger than 100 Myr in order to mitigate the effect of the stochastic spawning of star particles in hydrodynamical simulations.  Specifically, we replace such young star particles by drawing new particles with ages randomly assigned from a uniform distribution up to 100 Myr, positions randomly assigned within a kernel around the parent particle position, and masses drawn from the empirical power-law distribution of molecular cloud masses in the Milky Way \citep{Heyer2001},
\begin{equation}
    \frac{{\rm d}N}{{\rm d}M} \propto M^{-1.8}\ {\rm with}\ M \in [10^3,\ 10^6]\ M_{\odot},
\end{equation}
while ensuring mass conservation in the resampling process.

We further apply additional attenuation to a fraction of the stellar populations younger than 10Myr in order to mimic the attenuation from birth cloud dust (see R26). For accuracy, we run separate SKIRT radiative transfer calculations on birth clouds and build up a birth cloud obscured SED and dust re-emission library. We then treat those young star particles in birth clouds with the birth cloud SED templates.

Cameras are set up to observe galaxies under different viewing angles and at different wavelengths. They record the galaxy image at each wavelength in a field of view $20R_{\rm{star}}\times 20R_{\rm{star}}$. Mock images within observed-frame filters are obtained by redshifting the SKIRT datacube, applying attenuation by the intergalactic medium following \citet{Meiksin2006}, and convolving with the respective filter throughput curve.

\begin{table*}
\setlength{\tabcolsep}{2pt}
\centering 
\resizebox{\linewidth}{!}{
\begin{tabular}{c|c|c|c|c|c|c|c|c|c|c|c|c|c|c|c|c|c|c|c|c}
\hline \hline
 Facility & \multicolumn{5}{|c|}{{\it HST}/ACS} 
 & \multicolumn{7}{|c|}{{\it JWST}/NIRCam} 
 & {\it JWST}/MIRI
 & {\it Spitzer}/MIPS
 & \multicolumn{5}{|c|}{{\it Herschel}/PACS+SPIRE}
 & ALMA \\
 \cmidrule(lr){2-6} \cmidrule(lr){7-13} \cmidrule(lr){14-14} \cmidrule(lr){15-15} \cmidrule(lr){16-20} \cmidrule(lr){21-21} 
 Band & F435W & F606W & F775W & F814W & F850LP 
 & F090W & F115W & F150W & F200W & F277W
 & F356W & F444W & F1500W 
 & 24$\mu$m
 & 100$\mu$m & 160$\mu$m & 250$\mu$m & 350$\mu$m & 500$\mu$m
 & 870$\mu$m \\
\hline
$\lambda_{\rm{eff}}~[{\rm{\mu m}}]$ & 0.434 & 0.581 & 0.765 & 0.797 & 0.900 & 0.898 & 1.143 & 1.487 & 1.968 & 2.728 & 3.529 & 4.350 & 14.926 & 24 & 101 & 162 & 249 & 348 & 501 & 870 \\
\hline
IRres & \doublecheck & \doublecheck & \doublecheck & \doublecheck & \doublecheck & \doublecheck & \doublecheck & \doublecheck & \doublecheck & \doublecheck & \doublecheck & \doublecheck & \doublecheck & \checkmark & \checkmark & \checkmark & \checkmark & \checkmark & \checkmark & \doublecheck \\
IRint & \doublecheck & \doublecheck & \doublecheck & \doublecheck & \doublecheck & \doublecheck & \doublecheck & \doublecheck & \doublecheck & \doublecheck & \doublecheck & \doublecheck & \checkmark & \checkmark & \checkmark & \checkmark & \checkmark & \checkmark & \checkmark & \checkmark\\
noIR & \doublecheck & \doublecheck & \doublecheck & \doublecheck & \doublecheck & \doublecheck & \doublecheck & \doublecheck & \doublecheck & \doublecheck & \doublecheck & \doublecheck & \xmark & \xmark & \xmark & \xmark & \xmark & \xmark & \xmark & \xmark\\
nores & \checkmark & \checkmark & \checkmark & \checkmark & \checkmark & \checkmark & \checkmark & \checkmark & \checkmark & \checkmark & \checkmark & \checkmark & \checkmark & \checkmark & \checkmark & \checkmark & \checkmark & \checkmark & \checkmark & \checkmark \\
\hline
\hline
\end{tabular}
}
\caption{Introduction of four experiments with different (spatially-resolved) photometric wavelength coverage of the observational data. Single (\checkmark) and double (\doublecheck) checkmarks indicate the availability of galaxy-integrated and resolved information, respectively, whereas crosses ($\times$) denote no information for the respective band is included. The `IRres' run assumes available panchromatic and spatially resolved (with the exception of {\it Spitzer} and {\it Herschel}) data. `IRint' indicates a lack of spatial information at wavelengths longer than F444W. `noIR' represents no observations beyond F444W. `nores' is the case where only integrated photometry is available, i.e., reducing the input information to that in standard SED fitting.
}
\label{tab:filters}
\end{table*}

\subsubsection{From mock images to spectral distributions}
\label{sec:mockim2SD}

The {\tt SE3D} methodology does not work on the images directly, but rather on one-dimensional spectral distributions (SDs) derived from them, capturing the flux, size, S\'{e}rsic index and projected axis ratio in different observed-frame wavebands.
The integrated flux in a given waveband is obtained by direct summation across the image.  For the structural parameters, we assume the projected axis ratio $q_{\rm light}$ to be a weighted sum of the projected axis ratio of the stellar and dust distribution:
\begin{equation}
q_{\rm{light}}=f_{\rm{star}}q_{\rm{star}}+(1-f_{\rm{star}})q_{\rm{dust}},
\end{equation}
where $f_{\rm{star}}$ represents the fraction of light directly contributed by stars (as opposed to dust re-emission) at the wavelength of interest.  $q_{\rm{star}}$ and $q_{\rm{dust}}$ encode the projected mass profile of stars and dust under a certain inclination $\theta$:
\begin{equation}
    \begin{aligned}
     q_{\rm{star}}^2 &=\cos^2(\theta)+(C_{\rm{star}}/R_{\rm{star}})^2 \sin^2(\theta) \\
     q_{\rm{dust}}^2 &=\cos^2(\theta)+(C_{\rm{dust}}/R_{\rm{dust}})^2 \sin^2(\theta)  \\
    \end{aligned}  
\end{equation}
We then measure the half-light radius, $R_{\rm light}$, and the S\'{e}rsic index best describing the light profile, $n_{\rm light}$, using the curve of growth (i.e., cumulative light distribution) extracted from the image within elliptical annuli of shape $q_{\rm light}$.

We note that, in principle, one could take the creation of mock observations yet a step further by generating from the RT output synthetic images with survey realism (i.e., applying appropriate PSF convolution and noise levels), and conducting an end-to-end recovery exercise starting from these synthetic images, using the same photometric extraction tools and 2D surface brightness fitting software as observers do.  However, extensive tests on reliable extraction of photometry and global structural parameters from imaging at different depths and spatial resolution have been presented in the literature before \citep[e.g.,][]{Haussler2007, Haussler2013, Merlin2015, Wright2016, Weaver2023}.  Rather than testing these aspects of recovery explicitly, we will therefore assume that bias-free photometry and structural parameters are retrieved, and simply apply fiducial uncertainties by randomly perturbing the SDs assuming Gaussian error distributions (see Section\ \ref{sec:recov_toy} for details).

\subsection{SE3D fitting framework}
\label{sec:framework}

For each mock-observed galaxy, we run {\tt SE3D} fitting on the following four spectral distributions: the galaxy's integrated SED, and three Spectral Structural Distributions capturing respectively the wavelength-dependent size, S\'{e}rsic index and projected axis ratio. By default, we assume multi-wavelength observations to exist in a standard set of filters (see Table\ \ref{tab:filters}) from {\it HST}/ACS, {\it JWST}/NIRCam+MIRI, {\it Spitzer}/MIPS, {\it Herschel}/PACS+SPIRE to ALMA, with measurements of global structural parameters being available in all but the {\it Spitzer} and {\it Herschel} bands.  In Section\ \ref{sec:recov_toy}, we will further consider the impact of reduced data availability, in the infrared and/or at a spatially resolved level.\footnote{While the default wavelength sampling adopted in our analysis represents that of a well-studied deep field, it is by no means atypical for today's legacy lookback surveys.  Indeed, legacy surveys in fields such as COSMOS and GOODS-North/South have accumulated photometry in over 30 bands.  There is no inherent limitation to leverage such information using the {\tt SE3D} approach (i.e., by design a different filter set does not require retraining the emulator; see R26).  The choice of a more restricted 'standard' filter set is motivated primarily by the desire to not only represent the best case scenario.}

In order to infer constraints on the galaxy's physical properties, we use the affine invariant Markov chain Monte Carlo (MCMC) ensemble sampler {\tt emcee} by \citet{Foreman-Mackey2013}, which -in each realization of the model function- calls the ML emulator to predict the spectral distributions.  For a schematic overview of the procedure, we refer the reader to Figure 1 of R26.  While fitting, we adopt flat priors on all parameters within the bounds specified in R26's Table 1, except for stellar metallicity (where we adopt the normal distribution centred on Solar specified in R26's Table 1), and stellar and dust disk thickness (for which we adopt as prior a truncated Gaussian distribution with mean and standard deviation $(\mu,\ \sigma) = (0.2,\ 0.1)$ and bounds $[0.1,\ 1]$).  In practice, experiments with adjusted priors did not significantly alter the conclusions of our analysis.  In all fitting, we fix the age since the onset of star formation (parameter Age) to the age of the Universe at the galaxy redshift, which is assumed to be known.  Mock observations for TNG50 were generated and fitted with a fixed birth cloud covering fraction of young ($< 10$ Myr) stars, namely $f_{\rm cov} = 0.5$ (see R26 for details).  In the toy model case, the full range of $0 < f_{\rm cov} < 1$ was adopted in generating the mock observations and allowed in the fitting.

\subsection{Defining TNG "truth" in a toy model context}
\label{sec:true}

In order to compare physical properties estimated via parametrized modelling to TNG ground truth, we need to clarify how the latter is established.  For some quantities (e.g., $M_{\rm star}$ and $M_{\rm dust}$), this definition is trivial and can be directly extracted from the TNG particle information.  In other cases, however, it requires describing a TNG galaxy in a more simplified form.

First, we approximate their shape as an axisymmetric structure, the orientation of which is determined on the basis of the stellar angular momentum vector, aligned with the $z$-axis.  To determine the stellar angular momentum vector, we consider the dynamics of star particles within a radius $R$ in the interval $[R_{\rm half},\ {\rm max(30\ kpc,\ 3R_{half})]}$ that maximizes the fraction of kinetic energy in rotational motion ($f_{\rm rot} = \sum m_i\ v_{\phi,i}^2 / \sum m_i\ v_i^2$).  Here, $R_{\rm half}$ is the SubhaloHalfmassRad parameter in the TNG subhalo catalog, corresponding to the radius of a 3D sphere enclosing half of the subhalo's stellar mass.  

Having established the orientation of the galaxy, we can now meaningfully define the viewing angle $\theta$ for any camera, and measure the stellar half-mass radius $R_{\rm star}$ as the radius of a circle that encompasses half the stellar mass in the $xy$ plane.  The dust radius $R_{\rm dust}$ is measured equivalently from the projected dust mass distribution under the same orientation.  Averaging the projected stellar mass distribution in the $xz$ and $yz$ planes, we next define the height of the stellar distribution, $C_{\rm star}$, as the semi-minor axis length of an ellipse (with semi-major axis length fixed to $R_{\rm star}$) that encompasses half the mass.  Again, $C_{\rm dust}$ is quantified equivalently.

To determine the S\'{e}rsic index that best describes the stellar (or dust) mass profile, we construct the normalized cumulative projected mass distribution in the face-on view (i.e., $xy$ plane).  Via least-squares minimization we next determine which S\'{e}rsic index yields the smallest residuals at the radii encompassing 10\%, 20\%, 30\%, ..., 70\% of the mass.  In this fitting process, $n_{\rm star}$ (or $n_{\rm dust}$) is the only free parameter, as we keep $R_{\rm star}$ ($R_{\rm dust}$) fixed at the actual half-mass radius.  The motivation not to include points on the curve of growth beyond 70\% stems from the fact that TNG mass profiles are not always well approximated by a S\'{e}rsic profile (i.e., a form of model mismatch; see Section\ \ref{sec:modelmismatch}).  Including the outermost points on the curve of growth would yield a poorer approximation of the inner profile, where observationally most constraints would come from.  For a more in depth discussion on galaxy outskirts in TNG and a comparison to observed galaxies, we refer the reader to \citet{Merritt2020}.

The toy model galaxy description employed in {\tt SE3D} fitting assumes a lognormal SFH at each radius, characterized by the time of peak star formation, $t_{\rm peak}$, and the full-width-at-half-maximum, fwhm.  As a function of radius, the values of $\log(t_{\rm peak})$ and $\log({\rm fwhm})$ are allowed to vary linearly, captured by the gradients $k_{\rm peak}$ and $k_{\rm fwhm}$ (see R26 for details).  With the gradients and the values of SFH parameters at $R_{\rm star}$ (i.e., $t_{\rm peak,R_{star}}$ and ${\rm fwhm_{R_{star}}}$) in hand, the following implied properties can be computed:
\begin{itemize}
\item[\textbullet] the total star formation rate, SFR
\item[\textbullet] the half-SFR radius, $R_{\rm SFR}$
\item[\textbullet] the mass-weighted stellar age, ${\rm age_w}$
\item[\textbullet] the age gradient,\footnote{While technically of units Gyr, and not Gyr/$R_{\rm star}$ or Gyr/dex, we will for simplicity use the short-hand "age gradient" to refer to this derived quantity.} defined as the difference between the mass-weighted age outside versus inside $R_{\rm star}$: $\nabla {\rm age_w} \equiv {\rm age_w}(>R_{\rm star}) - {\rm age_w}(<R_{\rm star})$
\end{itemize}

\begin{figure}
    \centering
    \includegraphics[width=\linewidth]{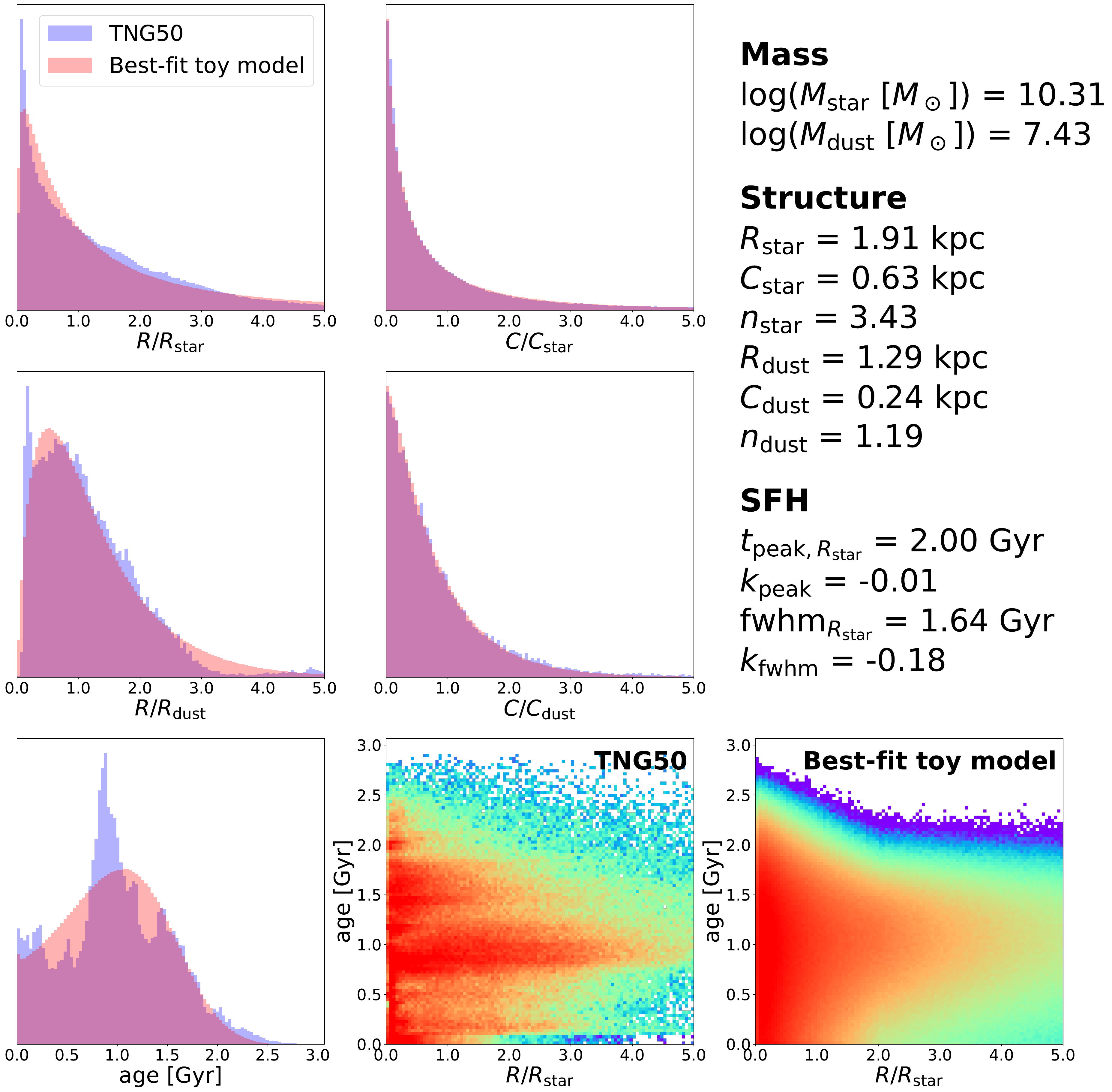}
    \caption{Structural and SFH properties of an example TNG galaxy contrasted with the closest toy model approximation.  The PDF of the radial and vertical stellar mass and dust distribution as well as the stellar age distribution are shown in blue for TNG, and in red for the associated toy model approximation.  The distribution of stellar mass across the age-radius plane (bottom-middle panel) is contrasted to that for the closest parametrized toy model (bottom-right panel).  Derived properties of the TNG galaxy are listed in the right space.}
    \label{fig:TNG_true_params}
\end{figure}

Unlike the toy model parameters $t_{\rm peak,R_{star}}$, ${\rm fwhm_{R_{star}}}$, $k_{\rm peak}$ and $k_{\rm fwhm}$, counterparts to these derived quantities can be measured directly for a TNG galaxy on the basis of its particle information.  We will therefore focus on these derived quantities when comparing recovered properties to TNG ground truth.  Nevertheless, to illustrate that also in the (resolved) SFH there is a form of template mismatch between the toy model description and actual TNG galaxies, we depict in Figure\ \ref{fig:TNG_true_params} for a case example the integrated and resolved SFH for both.  Projected spatial distributions of stars and dust are also compared in the radial and vertical dimension.  Altogether, the global structural and stellar population characteristics of the example galaxy are reasonably captured by the toy model description, although in a smoothed manner.

\section{Results}
\label{sec:results}

We present the results of our analysis in three steps.  First, we address key intrinsic and observed properties of the TNG galaxies in our sample, and contrast the latter to toy model galaxies in our library as well as real galaxies (Section\ \ref{sec:prop}).  Subsequently, we cover the recovery tests on mock-observed toy models and simulated galaxies (Sections\ \ref{sec:recov_toy} and\ \ref{sec:recov_TNG}, respectively).

\subsection{Intrinsic and observed properties of TNG galaxies}
\label{sec:prop}

\begin{figure}
    \centering
    \includegraphics[width=\linewidth]{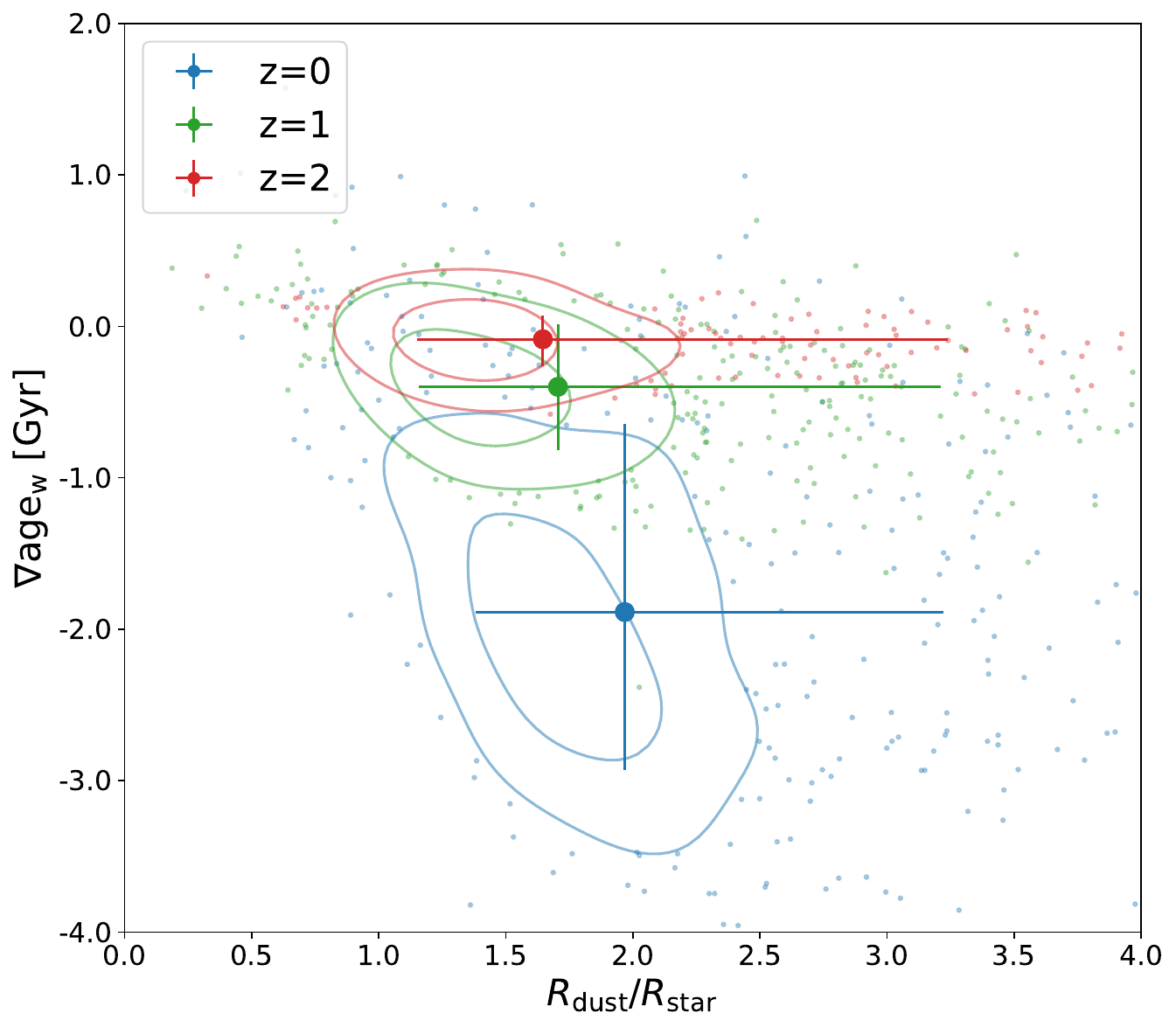}
    \includegraphics[width=\linewidth]{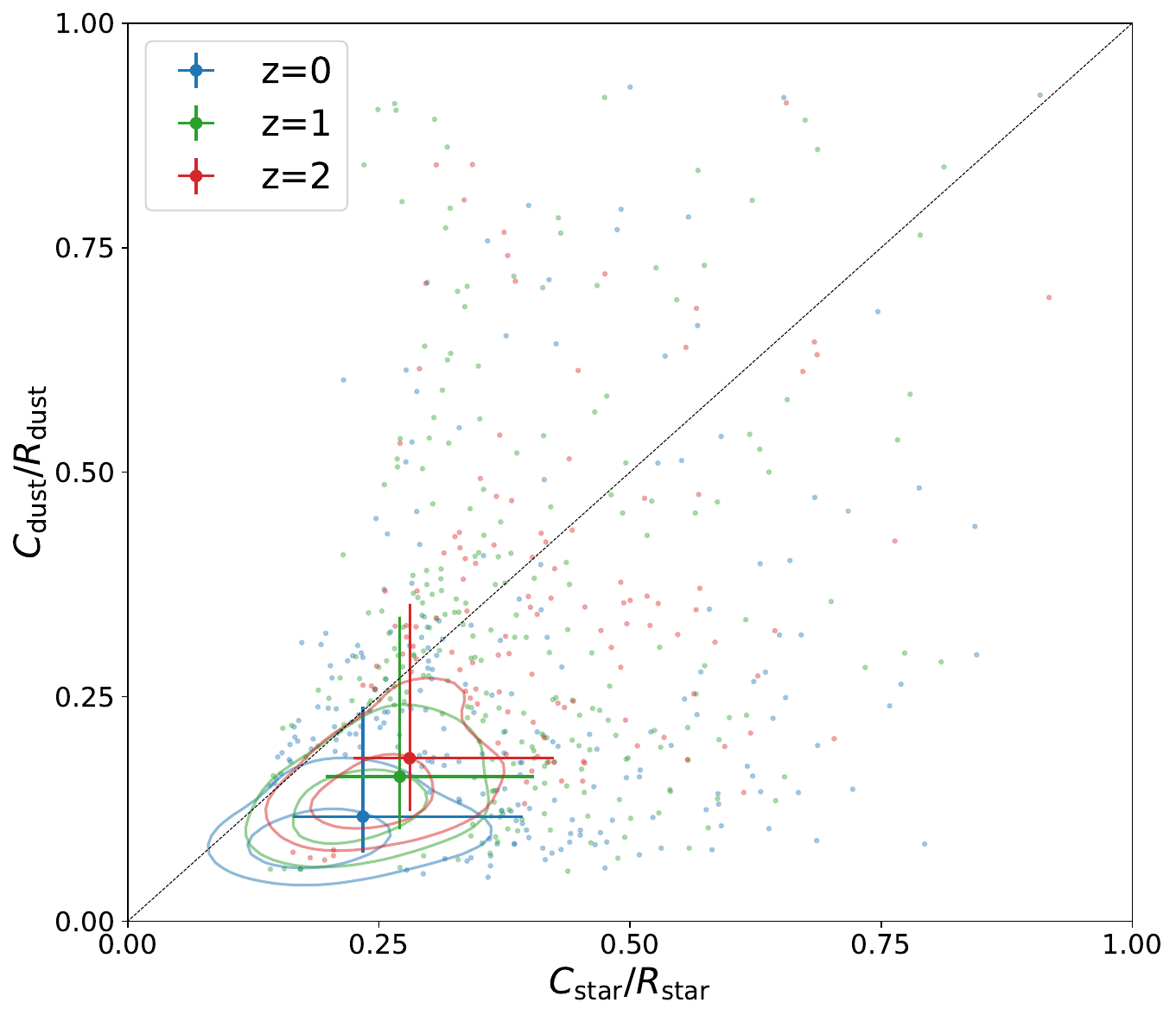}
    \caption{{\it Top}: Age gradient $\nabla {\rm{age}}_{\rm{w}}$ versus the relative radial extent of dust and stellar distributions, $R_{\rm{dust}}/R_{\rm{star}}$, of TNG50 galaxies at redshifts $z = 0$, 1 and 2. {\it Bottom}: Geometric thickness of the dust versus stellar distribution.  In each panel and for each redshift, the filled circle and cross mark the median and central 68th percentile of the distribution, respectively.}
    \label{fig:TNG_params_dist}
\end{figure}

\begin{figure*}
    \centering
    \includegraphics[width=\linewidth]{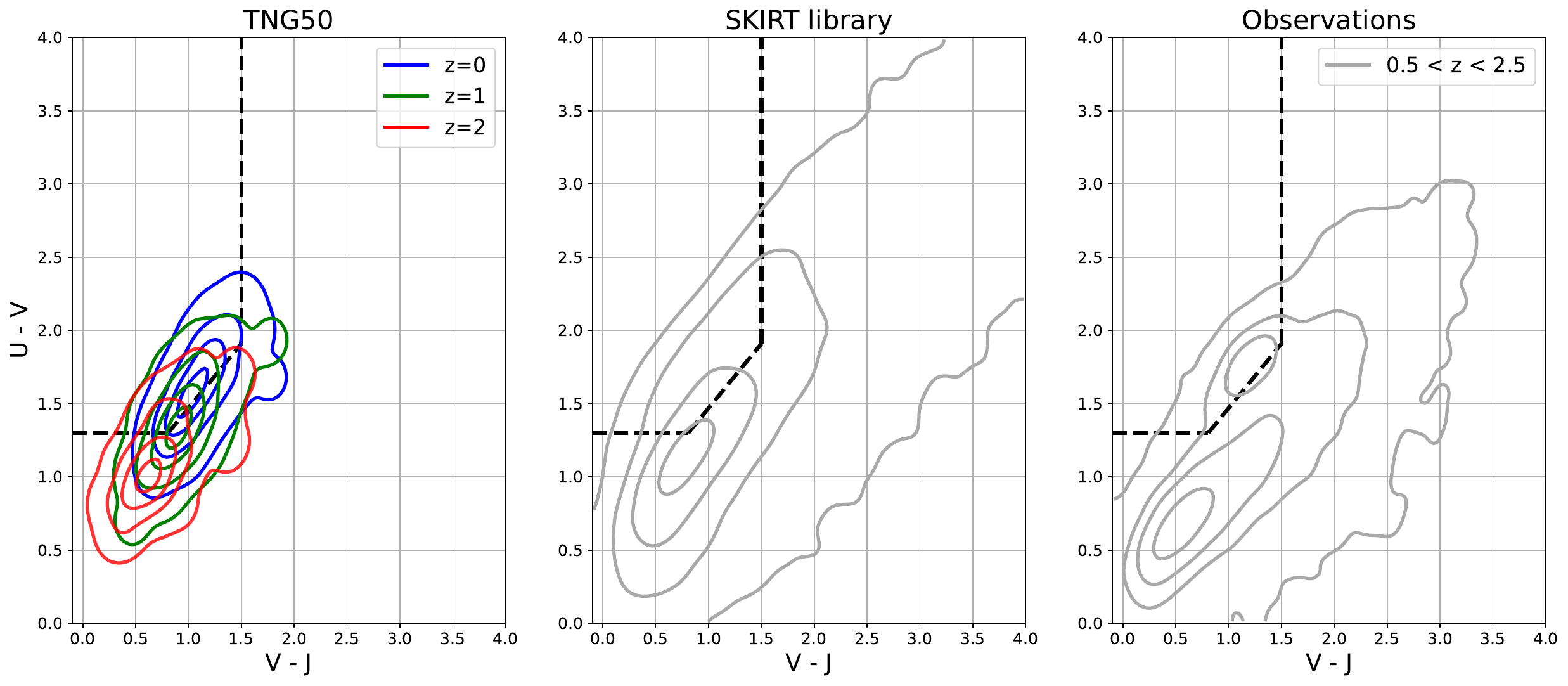}
    \caption{$UVJ$ diagram for massive SFGs in TNG50 ({\it left}), for the full set of toy model galaxies in our SKIRT library ({\it middle}), and for a mass-complete sample of observed galaxies at $0.5 < z < 2.5$ ({\it right}).  Contours depict the regions enclosing 20\%, 60\%, 90\% and 99.9\% of the data.  The colours predicted for TNG galaxies get bluer with increasing redshift and do not extend into the red $V-J$ regime.}
    \label{fig:UVJ_TNG_SKIRT_OBS}
\end{figure*}

\subsubsection{Resolved stellar and dust properties}
\label{sec:intrinsic}

In Figure\ \ref{fig:TNG_params_dist}, we show that massive SFGs in TNG span a range in resolved properties, with systematic shifts between the distributions from $z=2$ to $z=0$.  At cosmic noon ($z=2$), stellar age gradients are modest, albeit predominantly (for 74.39\% of the galaxies) negative.  Over time, these then develop from a median difference between outer and inner stellar ages of -0.09 Gyr ($z=2$) to -0.42 Gyr ($z=1$), and ultimately -1.91 Gyr ($z=0$).  Of course, the dividing line between what constitutes outer and inner stars (i.e., $R_{\rm star}$) also increases in an absolute (kpc) sense over this timespan, and moreover the samples of mass-selected SFGs at the different redshifts only partially include each other's progenitors/descendants.  Nevertheless, the redshift trend is in line with an inside-out evolutionary picture.  In units normalized to the age of the Universe, the age gradients at different redshifts are more similar, but still change systematically from $\frac{\nabla {\rm age_w}}{\rm age_{Universe}} = -0.03$ at $z=2$ to -0.07 at $z=1$ and ultimately -0.14 at $z=0$.

At all epochs considered, the vast majority of SFGs feature dust disks that are radially more extended than the stellar distribution.  This is especially the case at later times, with the median $R_{\rm dust}/R_{\rm star}$ increasing from 1.63 to 1.97 between cosmic noon and the present day.  This too may be considered an imprint of inside-out growth, with the dust sampling the cold ISM out of which new stars form.  However, it is not trivial either, as negative metallicity gradients -particularly present at late times- reduce the dust-to-gas ratio in galaxy outskirts.  At first glance, it also appears at odds with observational findings of compact dust cores in cosmic noon galaxies \citep[e.g.,][]{Hodge2016,Tadaki2017b,Tadaki2020,Puglisi2019,Martorano2025}. No strong correlation between $\nabla {\rm age_w}$ and $R_{\rm dust}/R_{\rm star}$ is observed.  We note that the extremes in $R_{\rm dust}/R_{\rm star}$, both on the $R_{\rm dust} < R_{\rm star}$ side and for $R_{\rm dust} \gg R_{\rm star}$, are found among SFGs with relatively flat age gradients.  As documented in Appendix\ \ref{app:dustassignment}, these results are only modestly affected by the detailed approach of how dust is `painted' onto TNG galaxies in post-processing.  That said, it is useful to bear in mind that these results do not necessarily extrapolate to and would in fact be worthwhile to revisit for models that explicitly incorporate the physics of dust formation and destruction \citep[e.g.,][]{Vijayan2019,Li2019,Trayford2026}.  Finally, we note that the toy model galaxies in our library span a range $0.1 < R_{\rm dust} / R_{\rm star} < 4$ and $-5.21 < \nabla {\rm age_w} < 4.89$, encompassing all but a few of the massive SFGs in TNG. 

Turning to the thickness of the stellar and dust distributions, we find systematic changes with redshift here too.  At all epochs, the dust disks are generally flatter (i.e., have smaller $C/R$) than the stellar distribution, by on average a factor $\sim 0.66$.  From cosmic noon to the present day, the median $C_{\rm star}/R_{\rm star}$ decreases from 0.28 ($z=2$) to 0.27 ($z=1$) and ultimately 0.23 ($z=0$).  For $C_{\rm dust}/R_{\rm dust}$ the equivalent evolution ranges from 0.18 over 0.16 to 0.12.  Of course, these statistics are only descriptive for the bulk of the population and a tail towards fairly round shapes, both in dust and in stars, is notable as well.  The geometric thickness quantified in Figure\ \ref{fig:TNG_params_dist} is reflected in the galaxies' dynamical properties too.  We verified that the geometric thickness $C/R$ of each component (stars/dust) correlates strongly with the fraction of kinetic energy in circular rotational motion (i.e., along the $\phi$ dimension of a cylindrical coordinate system) for that component.  While the observed trend is in line with a picture where high-z disks are more turbulent and both dynamically and geometrically thicker \citep[see, e.g.,][]{Wisnioski2019, Ubler2019,Tiley2021}, we caution that numerical effects may also contribute.  High-z galaxies are on average smaller and their scaleheight in absolute (kpc) units is therefore closer to the resolution limit.  While direct resolution limitations are mitigated by focussing on massive ($> 10^{10}\ M_{\odot}$) SFGs, the lack of thin stellar structures in cosmologically simulated galaxies has been attributed to indirect resolution-related numerical effects. Specifically, the relatively high mass of dark matter particles relative to baryonic particles can induce spurious heating, leaving stellar structures artificially inflated \citep[e.g.,][]{Ludlow2021, deGraaff2022, Zhang2022}.

We conclude that our sample of massive SFGs spans a significant range in parameters that specify their intrinsic stellar and dust make-up.

\begin{figure}
    \centering
    \includegraphics[width=\linewidth]{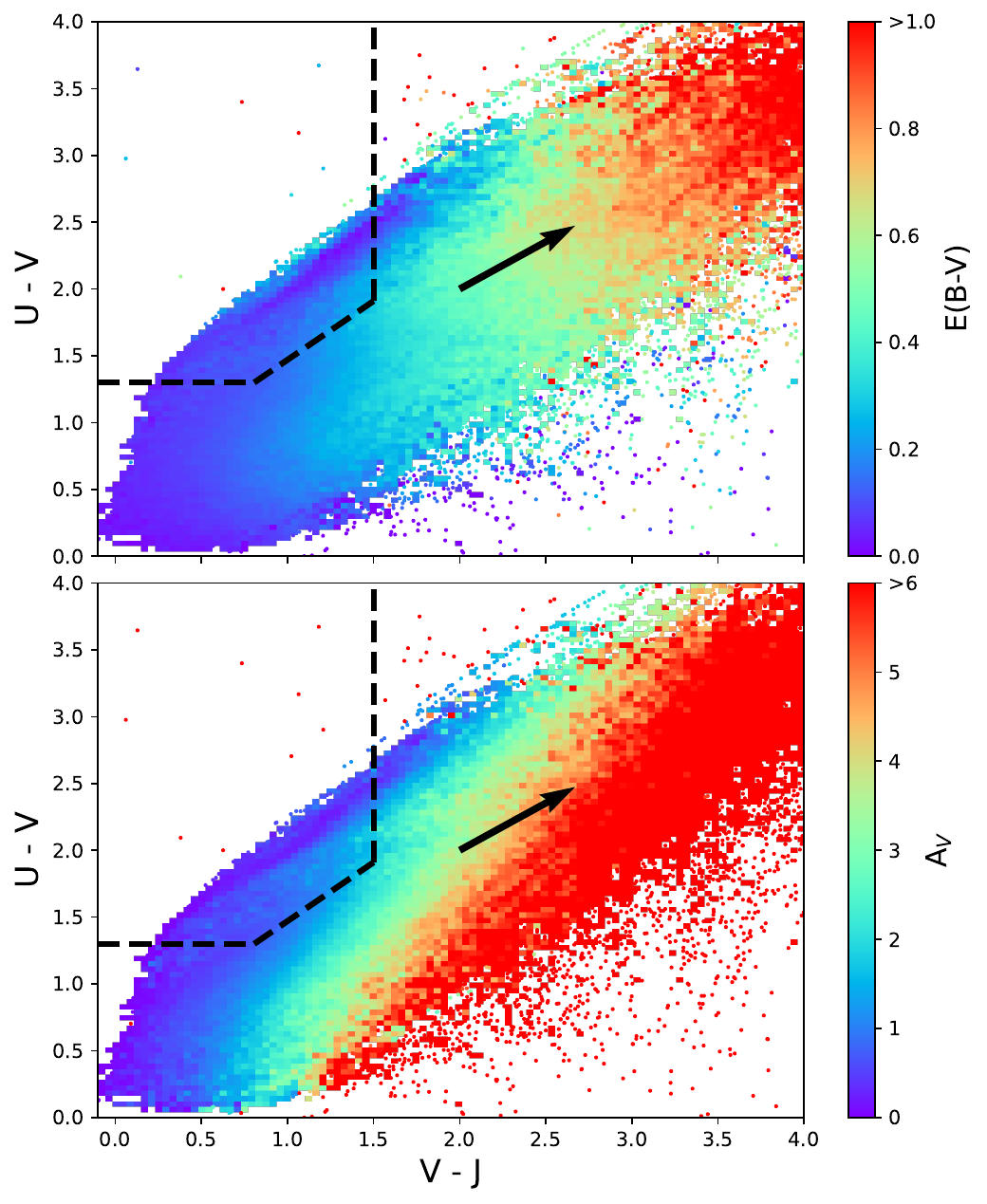}
    \caption{$UVJ$ diagram for our toy model library, colour-coded by the effective reddening, $E(B-V)$, and the effective visual attenuation, $A_V$. For reference, the black arrow depicts the dust reddening (or equivalently attenuation) vector computed using the \citet{Calzetti2000} attenuation law, for $A_V$ = 1. For our toy models of mixed stars and dust, reddening and attenuation increase along different directions in the $UVJ$ plane.}
    \label{fig:UVJ_EBV_Av}
\end{figure}

\begin{figure*}
    \centering
    \includegraphics[width=\linewidth]{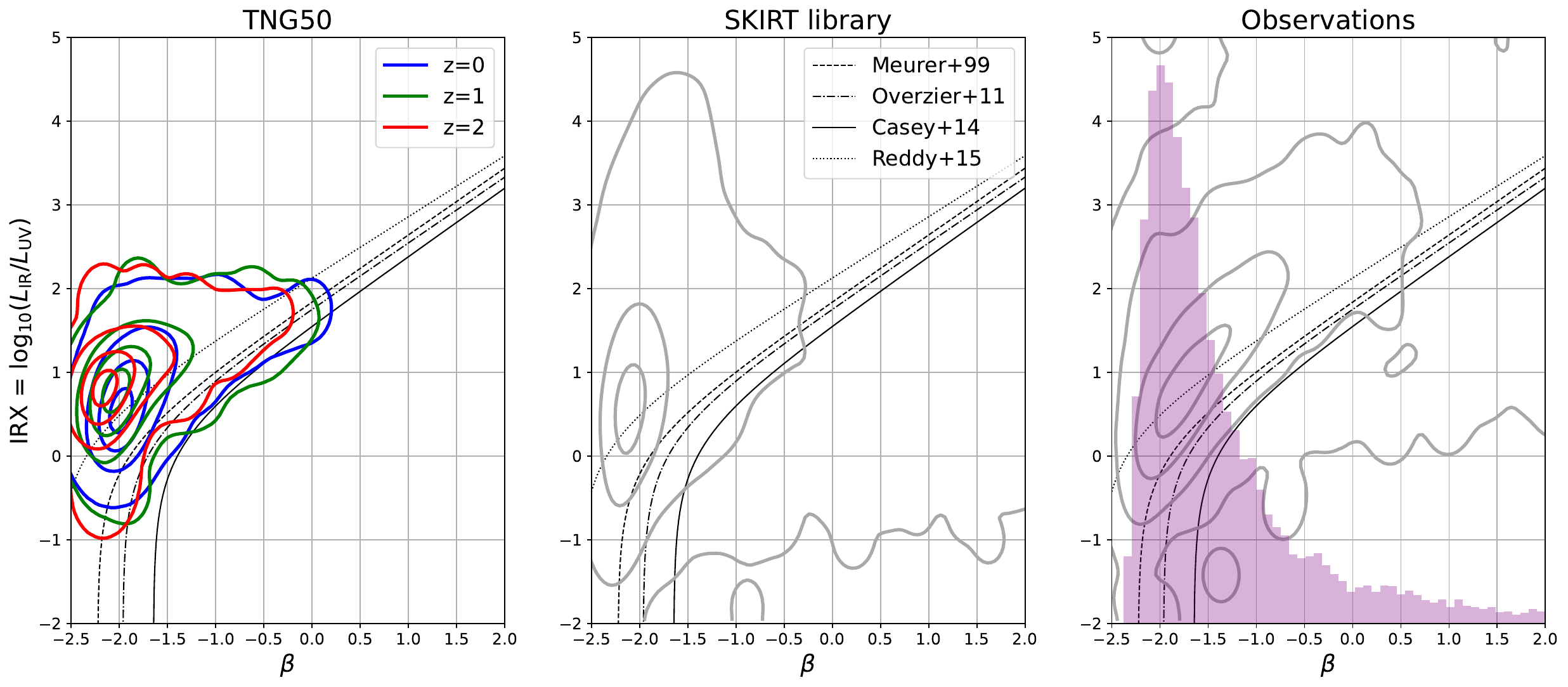}
    \caption{IRX - $\rm{\beta}$ diagram for massive SFGs in TNG50 ({\it left}), for the full set of toy model galaxies in our SKIRT library ({\it middle}), and for a mass-limited sample of observed galaxies at $0.5 < z < 2.5$ ({\it right}).  Contours depict the regions enclosing 20\%, 60\%, 90\% and 99.9\% of the data. On the right panel, we overlay the distribution of available $\beta$ from observations, whilst the IRX - $\rm{\beta}$ contours are only shown for galaxies with available IRX measurements.}
    \label{fig:IRXbeta}
\end{figure*}

\subsubsection{Observational diagnostic diagrams}
\label{sec:obsdiagrams}

We now turn to what the selected TNG galaxies look like in observed space.  Specifically, we consider two rest-frame diagnostic diagrams that are often utilized to trace the impact of dust across observed (or simulated) galaxy populations: the $UVJ$ and IRX-$\beta$ diagrams.

In Figure\ \ref{fig:UVJ_TNG_SKIRT_OBS}, we show the $UVJ$ colour distribution of our sample of massive SFGs extracted from TNG50 (left panel), alongside that of our full SKIRT library of toy model galaxies (middle panel, see R26 for details), and a mass-complete sample of observed galaxies since cosmic noon (right panel).  We stress that no effort is made to match the samples by either redshift, stellar mass or star formation activity.  The middle and right panels thus only serve for reference.  Specifically, the TNG galaxies shown are selected from three discrete redshifts, required to be star-forming on the basis of their sSFR, and more massive than $10^{10}\ M_{\odot}$.  The SKIRT library on the other hand extends down to $10^{9.5}\ M_{\odot}$, considered a continuous distribution of times since the onset of star formation in the interval $\log({\rm{Age}\ [Gyr]}) \in [0.0, 1.1]$, and contains a broad range of star-formation activities, including quiescent objects.  We further note that, deliberately, no covariances between their SFH and structural properties were introduced.  Finally, the sample of 17,207 observed galaxies depicted in the rightmost panel covers the redshift range $0.5 < z < 2.5$ (avoiding potential issues with aperture corrections for the largest low-z objects), and were selected to have a stellar mass $\log(M_{\rm star}) > 9.5$ without further consideration of their star formation activity (hence the prominent clump of objects in the quiescent wedge).  They were extracted from the DAWN JWST Archive (DJA) morpho-photometric catalogue combining the GOODS-N/S, CEERS, PRIMER-UDS and PRIMER-COSMOS JWST survey fields \citep{Genin2025}.\footnote{\href{https://dawn-cph.github.io/dja/blog/2024/08/16/morphological-data/}{https://dawn-cph.github.io/dja/blog/2024/08/16/morphological-data/}}

Sample differences withstanding, the following conclusions can be drawn.  First, TNG galaxies do not span as broad a colour range as those observed, nor when compared to our library of toy model galaxies.  Especially in $V-J$ the observed and toy model galaxies reach redder colours \citep[see also][]{Gebek2025}.  Imposing additional criteria to eliminate from the SKIRT library or the DJA sample objects of quiescent nature or with stellar mass below $10^{10}\ M_{\odot}$ would not alleviate this discrepancy, as the reddest objects in $V-J$ tend to be massive and star-forming.  Secondly, the colours of massive SFGs in TNG extend into the quiescent wedge of the $UVJ$ diagram (demarcated by the dashed black lines), especially at low redshift.  While cases where the $UVJ$- and sSFR-based classifications of quiescence differ exist among observed galaxies, such classification discrepancies generally remain well below 10\% in observed samples \citep{Lang2014, Zhang2022}. Thirdly, across the redshift interval considered, the mean colour evolution of TNG SFGs is one where both $U-V$ and $V-J$ colours get bluer with increasing redshift.  We attribute this primarily to the stellar populations' younger ages causing bluer intrinsic emission, paired with a lack of evolution in $M_{\rm dust}/M_{\rm star}$ to offset this effect via dust reddening.  Evolution in stellar metallicities and star-dust geometries can influence the relative colour distributions too\footnote{Stellar metallicity affects primarily the intrinsic $V-J$ colour, much less so $U-V$, according to the sensitivity analysis presented in R26's Figure 8.}, but are likely subdominant effects.  We discuss the realism of mock-observed TNG galaxies in more depth in Section\ \ref{sec:obssim}.

Regarding the impact of dust on $UVJ$ colours, the diagonal locus of SFGs is normally interpreted as a sequence of increasing dust reddening, due to enhanced dust content and/or (folding in geometry such as galaxy size and viewing angle) enhanced dust columns \citep[e.g.,][]{Wuyts2007, Williams2009, Patel2012, Zuckerman2021}.  Under the assumption of a uniform foreground screen of dust, this dust reddening is then equated to a fixed fraction of the total visual attenuation, i.e., $E(B-V) = \frac{A_V}{R_V}$.  As our toy model library features a broad range of both dust content and geometries, it is interesting to evaluate how the resulting effective reddening, $E(B-V) \equiv (B-V)_{\rm with\ dust} - (B-V)_{\rm no\ dust}$, and effective visual attenuation, $A_V \equiv A_{V, {\rm with\ dust}} - A_{V, {\rm no\ dust}}$, vary across the $UVJ$ plane.  This is illustrated in Figure\ \ref{fig:UVJ_EBV_Av}.  One can appreciate that dust reddening indeed increases along the diagonal sequence of SFGs, as anticipated:
\begin{equation}
{\rm Increasing}\ E(B-V): \frac{{\rm d}(U-V)}{{\rm d}(V-J)} = 1.51
\end{equation}
However, the net visual attenuation increases along a vector of different orientation:
\begin{equation}
{\rm Increasing}\ A_V: \frac{{\rm d}(U-V)}{{\rm d}(V-J)} = -0.31
\end{equation}
This implies that some of the most heavily attenuated objects exhibit moderately red $V-J$ yet relatively blue $U-V$ colours.  This is reminiscent of the ``skin effect'' discussed in the interpretation of IRX-$\beta$ diagrams \citep{Popping2017,Wang2018}, whereby the spectral slope at shorter wavelengths is more sensitive to the optically thin regions near the surface of galactic disks.  In our toy model setup, radially varying dust columns (and stellar populations) can further enhance this effect.  In other words, the toy model galaxies not only span a broad range in $E(B-V)$ and $A_V$, but also in $R_V$.  For a sensitivity analysis of which physical conditions impact the total-to-selective attenuation most, we refer the reader to R26 (Section 4).\footnote{The projected dust column, $\Sigma_{\rm proj, dust} \equiv \frac{0.5 M_{\rm dust}}{\pi R_{\rm dust}^2 q_{\rm dust}}$, is the main culprit.}

Turning to the IRX-$\beta$ diagram, we show in Figure\ \ref{fig:IRXbeta} the infrared excess, $IRX \equiv \log(L_{\rm IR}/L_{\rm UV})$, as a function of the UV slope, $\beta$, for the massive SFG sample in TNG50, for our full toy model library, and for the subset of IR-detected galaxies in the observed DJA sample.  Here, we define $L_{\rm IR} \equiv L(8 - 1000\mu{\rm m})$, $L_{\rm UV} \equiv \lambda\ L_{\lambda}(1600$\AA$)$, and quantify $\beta$ as the best-fitting power-law slope to the SED in the UV windows defined by \citet{Calzetti1994}.  Since the coverage and depth of mid- and far-IR surveys \citep{LeFloch2009,Lutz2011,Elbaz2011,Magnelli2013} is such that only a fraction of DJA galaxies have direct constraints on $L_{\rm IR}$, we also contrast for reference the density distribution of $\beta$ values for the full underlying DJA sample in the right-hand panel of Figure\ \ref{fig:IRXbeta}.  Finally, we show for context a number of IRX-$\beta$ relations from the literature \citep{Meurer1999, Overzier2011, Casey2014, Reddy2015}.

An immediate takeaway from Figure\ \ref{fig:IRXbeta} is that TNG galaxies exhibit a smaller spread than observed galaxies in IRX -- $\beta$ space.  This is the case along each of the two dimensions: both in IR excess and in UV slope.  Our SKIRT library of toy model galaxies comparatively spans a broader range of IRX and $\beta$ values, akin to what was also seen in the $UVJ$ diagram.  This suggests that the library encompasses toy models which, on account of their dust content and star-dust geometry, may be more comparable to some dusty SFGs in the real Universe than any cosmologically simulated galaxies are.  That said, the mode of the $\beta$ distribution for both toy model and simulated galaxies appears to be shifted towards slightly bluer UV slopes than seen in the observations.  From $z=0$ to $z=2$, the sample of massive SFGs in TNG shows only modest evolution, with the peak of the density distribution shifting towards higher IRX, yet bluer $\beta$ with increasing redshift.  This too echoes some of the trends observed in $UVJ$ space, and can likely be attributed to the same lack of strong evolution in dust content, paired with the aforementioned skin effect yielding an inefficient reddening. For a more in depth discussion of the IRX-$\beta$ distribution of TNG50 SFGs, including an evaluation of the impact of different dust types, we refer the reader to \citet{Schulz2020}.

We conclude that both in $UVJ$ and in IRX-$\beta$, the RT post-processing of TNG50 SFGs does not yield coverage across the full dynamic range exhibited by observed galaxies.  While not emerging from a self-consistent simulation and oversimplified in their structural and SFH make-up compared to both real and simulated galaxies, our toy model galaxies do cover (and indeed extend beyond) the dynamic range of real galaxies.  We therefore consider it worthwhile to run {\tt SE3D} fitting on both toy model and simulated galaxies, and carry out recovery tests of intrinsic properties on both types of synthetic datasets.  This is what we do in Sections\ \ref{sec:recov_toy} and\ \ref{sec:recov_TNG}, respectively.

\begin{table}
\centering 
\resizebox{\linewidth}{!}{
\begin{tabular}{c|c|c|c|c|c}
\hline \hline

Toy model & Redshift & IRres & IRint & noIR & nores \\
\hline

$\log(M_{\rm{star}})$ & \textcolor{blue}{0} & \textcolor{blue}{0.020$\pm$0.104} & \textcolor{blue}{0.024$\pm$0.118} & \textcolor{blue}{0.034$\pm$0.180} & \textcolor{blue}{0.006$\pm$0.149} \\
$[\log(M_\odot)]$ & \textcolor{green!50!black}{1} & \textcolor{green!50!black}{-0.002$\pm$0.099} & \textcolor{green!50!black}{0.005$\pm$0.112} & \textcolor{green!50!black}{0.007$\pm$0.210} & \textcolor{green!50!black}{-0.017$\pm$0.122} \\
 & \textcolor{red}{2} & \textcolor{red}{-0.010$\pm$0.085} & \textcolor{red}{0.005$\pm$0.097} & \textcolor{red}{-0.005$\pm$0.222} & \textcolor{red}{-0.019$\pm$0.105} \\
\hline
$\log(R_{\rm{star}})$ & \textcolor{blue}{0} & \textcolor{blue}{-0.015$\pm$0.071} & \textcolor{blue}{-0.005$\pm$0.099} & \textcolor{blue}{-0.015$\pm$0.122} & \textcolor{blue}{-0.028$\pm$0.172} \\
$[\log({\rm{kpc}})]$ & \textcolor{green!50!black}{1} & \textcolor{green!50!black}{-0.016$\pm$0.066} & \textcolor{green!50!black}{-0.014$\pm$0.085} & \textcolor{green!50!black}{-0.040$\pm$0.142} & \textcolor{green!50!black}{-0.045$\pm$0.181} \\
 & \textcolor{red}{2} & \textcolor{red}{-0.002$\pm$0.065} & \textcolor{red}{0.002$\pm$0.083} & \textcolor{red}{-0.004$\pm$0.160} & \textcolor{red}{0.001$\pm$0.181} \\
\hline
$C_{\rm{star}}/R_{\rm{star}}$ & \textcolor{blue}{0} & \textcolor{blue}{0.003$\pm$0.063} & \textcolor{blue}{0.003$\pm$0.070} & \textcolor{blue}{0.008$\pm$0.076} & \textcolor{blue}{0.005$\pm$0.084} \\
 & \textcolor{green!50!black}{1} & \textcolor{green!50!black}{-0.003$\pm$0.062} & \textcolor{green!50!black}{0.004$\pm$0.075} & \textcolor{green!50!black}{0.011$\pm$0.081} & \textcolor{green!50!black}{0.007$\pm$0.091} \\
 & \textcolor{red}{2} & \textcolor{red}{-0.009$\pm$0.060} & \textcolor{red}{-0.012$\pm$0.090} & \textcolor{red}{-0.002$\pm$0.082} & \textcolor{red}{-0.001$\pm$0.092} \\
\hline
$\log(n_{\rm{star}})$ & \textcolor{blue}{0} & \textcolor{blue}{0.002$\pm$0.096} & \textcolor{blue}{0.007$\pm$0.128} & \textcolor{blue}{0.009$\pm$0.175} & \textcolor{blue}{-0.006$\pm$0.398} \\
$[{\rm{dex}}]$ & \textcolor{green!50!black}{1} & \textcolor{green!50!black}{0.017$\pm$0.092} & \textcolor{green!50!black}{0.011$\pm$0.103} & \textcolor{green!50!black}{0.034$\pm$0.194} & \textcolor{green!50!black}{0.107$\pm$0.412} \\
 & \textcolor{red}{2} & \textcolor{red}{0.001$\pm$0.053} & \textcolor{red}{-0.009$\pm$0.118} & \textcolor{red}{0.014$\pm$0.244} & \textcolor{red}{0.051$\pm$0.387} \\
\hline
$\log(M_{\rm{dust}})$ & \textcolor{blue}{0} & \textcolor{blue}{0.025$\pm$0.076} & \textcolor{blue}{0.031$\pm$0.086} & \textcolor{blue}{0.014$\pm$0.308} & \textcolor{blue}{0.040$\pm$0.107} \\
$[\log(M_\odot)]$ & \textcolor{green!50!black}{1} & \textcolor{green!50!black}{0.020$\pm$0.068} & \textcolor{green!50!black}{0.043$\pm$0.096} & \textcolor{green!50!black}{0.034$\pm$0.286} & \textcolor{green!50!black}{0.044$\pm$0.109} \\
 & \textcolor{red}{2} & \textcolor{red}{-0.005$\pm$0.081} & \textcolor{red}{0.017$\pm$0.099} & \textcolor{red}{0.005$\pm$0.422} & \textcolor{red}{0.014$\pm$0.160} \\
\hline
$\log(R_{\rm{dust}})$ & \textcolor{blue}{0} & \textcolor{blue}{0.008$\pm$0.076} & \textcolor{blue}{0.040$\pm$0.171} & \textcolor{blue}{0.044$\pm$0.211} & \textcolor{blue}{0.070$\pm$0.328} \\
$[\log({\rm{kpc}})]$ & \textcolor{green!50!black}{1} & \textcolor{green!50!black}{0.010$\pm$0.086} & \textcolor{green!50!black}{0.040$\pm$0.168} & \textcolor{green!50!black}{0.012$\pm$0.238} & \textcolor{green!50!black}{0.024$\pm$0.327} \\
 & \textcolor{red}{2} & \textcolor{red}{0.008$\pm$0.097} & \textcolor{red}{0.037$\pm$0.193} & \textcolor{red}{0.046$\pm$0.315} & \textcolor{red}{0.061$\pm$0.403} \\
\hline
$C_{\rm{dust}}/R_{\rm{dust}}$ & \textcolor{blue}{0} & \textcolor{blue}{0.002$\pm$0.065} & \textcolor{blue}{0.005$\pm$0.071} & \textcolor{blue}{0.004$\pm$0.076} & \textcolor{blue}{0.024$\pm$0.085} \\
 & \textcolor{green!50!black}{1} & \textcolor{green!50!black}{0.002$\pm$0.062} & \textcolor{green!50!black}{0.011$\pm$0.074} & \textcolor{green!50!black}{0.017$\pm$0.073} & \textcolor{green!50!black}{0.022$\pm$0.084} \\
 & \textcolor{red}{2} & \textcolor{red}{-0.003$\pm$0.060} & \textcolor{red}{0.008$\pm$0.075} & \textcolor{red}{0.016$\pm$0.080} & \textcolor{red}{0.009$\pm$0.082} \\
\hline
$\log(n_{\rm{dust}})$ & \textcolor{blue}{0} & \textcolor{blue}{-0.014$\pm$0.145} & \textcolor{blue}{0.012$\pm$0.309} & \textcolor{blue}{0.059$\pm$0.396} & \textcolor{blue}{0.112$\pm$0.522} \\
$[{\rm{dex}}]$ & \textcolor{green!50!black}{1} & \textcolor{green!50!black}{0.009$\pm$0.177} & \textcolor{green!50!black}{0.036$\pm$0.386} & \textcolor{green!50!black}{0.091$\pm$0.446} & \textcolor{green!50!black}{0.066$\pm$0.570} \\
 & \textcolor{red}{2} & \textcolor{red}{0.010$\pm$0.225} & \textcolor{red}{0.006$\pm$0.382} & \textcolor{red}{0.071$\pm$0.456} & \textcolor{red}{0.204$\pm$0.461} \\
\hline
$\log({\rm{SFR}})$ & \textcolor{blue}{0} & \textcolor{blue}{0.034$\pm$0.127} & \textcolor{blue}{0.038$\pm$0.142} & \textcolor{blue}{0.030$\pm$0.171} & \textcolor{blue}{0.046$\pm$0.147} \\
$[\log(M_\odot/{\rm{yr}})]$ & \textcolor{green!50!black}{1} & \textcolor{green!50!black}{-0.001$\pm$0.078} & \textcolor{green!50!black}{0.001$\pm$0.088} & \textcolor{green!50!black}{0.025$\pm$0.187} & \textcolor{green!50!black}{0.002$\pm$0.100} \\
 & \textcolor{red}{2} & \textcolor{red}{0.006$\pm$0.091} & \textcolor{red}{-0.000$\pm$0.096} & \textcolor{red}{-0.017$\pm$0.314} & \textcolor{red}{-0.001$\pm$0.086} \\
\hline
$\log(R_{\rm{SFR}})$ & \textcolor{blue}{0} & \textcolor{blue}{-0.025$\pm$0.089} & \textcolor{blue}{-0.020$\pm$0.094} & \textcolor{blue}{-0.023$\pm$0.114} & \textcolor{blue}{-0.066$\pm$0.225} \\
$[\log({\rm{kpc}})]$ & \textcolor{green!50!black}{1} & \textcolor{green!50!black}{-0.007$\pm$0.061} & \textcolor{green!50!black}{-0.012$\pm$0.062} & \textcolor{green!50!black}{-0.013$\pm$0.144} & \textcolor{green!50!black}{-0.057$\pm$0.221} \\
 & \textcolor{red}{2} & \textcolor{red}{-0.015$\pm$0.053} & \textcolor{red}{-0.005$\pm$0.113} & \textcolor{red}{0.012$\pm$0.183} & \textcolor{red}{-0.018$\pm$0.235} \\
\hline
${\rm{age}}_w$ & \textcolor{blue}{0} & \textcolor{blue}{-0.047$\pm$1.080} & \textcolor{blue}{0.064$\pm$1.056} & \textcolor{blue}{0.194$\pm$1.275} & \textcolor{blue}{0.168$\pm$1.192} \\
$[{\rm{Gyr}}]$ & \textcolor{green!50!black}{1} & \textcolor{green!50!black}{-0.011$\pm$0.413} & \textcolor{green!50!black}{-0.037$\pm$0.490} & \textcolor{green!50!black}{0.045$\pm$0.531} & \textcolor{green!50!black}{-0.034$\pm$0.419} \\
 & \textcolor{red}{2} & \textcolor{red}{-0.043$\pm$0.190} & \textcolor{red}{-0.028$\pm$0.208} & \textcolor{red}{0.028$\pm$0.378} & \textcolor{red}{-0.031$\pm$0.261} \\
\hline
$\nabla {\rm{age}}_{\rm{w}}$ & \textcolor{blue}{0} & \textcolor{blue}{0.002$\pm$0.911} & \textcolor{blue}{-0.065$\pm$1.226} & \textcolor{blue}{-0.010$\pm$1.008} & \textcolor{blue}{0.174$\pm$1.691} \\
$[{\rm{Gyr}}]$ & \textcolor{green!50!black}{1} & \textcolor{green!50!black}{-0.081$\pm$0.428} & \textcolor{green!50!black}{-0.065$\pm$0.477} & \textcolor{green!50!black}{-0.083$\pm$0.543} & \textcolor{green!50!black}{0.042$\pm$0.677} \\
 & \textcolor{red}{2} & \textcolor{red}{0.001$\pm$0.201} & \textcolor{red}{-0.012$\pm$0.302} & \textcolor{red}{-0.014$\pm$0.283} & \textcolor{red}{0.025$\pm$0.332} \\
\hline
$Z$ & \textcolor{blue}{0} & \textcolor{blue}{0.045$\pm$0.451} & \textcolor{blue}{0.072$\pm$0.485} & \textcolor{blue}{0.068$\pm$0.442} & \textcolor{blue}{0.004$\pm$0.431} \\
$[Z_\odot]$ & \textcolor{green!50!black}{1} & \textcolor{green!50!black}{0.103$\pm$0.410} & \textcolor{green!50!black}{0.060$\pm$0.445} & \textcolor{green!50!black}{0.092$\pm$0.408} & \textcolor{green!50!black}{0.093$\pm$0.447} \\
 & \textcolor{red}{2} & \textcolor{red}{0.078$\pm$0.474} & \textcolor{red}{0.040$\pm$0.437} & \textcolor{red}{0.093$\pm$0.485} & \textcolor{red}{0.086$\pm$0.428} \\
\hline
$f_{\rm{cov}}$ & \textcolor{blue}{0} & \textcolor{blue}{-0.001$\pm$0.189} & \textcolor{blue}{-0.020$\pm$0.186} & \textcolor{blue}{-0.028$\pm$0.294} & \textcolor{blue}{-0.017$\pm$0.232} \\
 & \textcolor{green!50!black}{1} & \textcolor{green!50!black}{-0.025$\pm$0.147} & \textcolor{green!50!black}{-0.021$\pm$0.164} & \textcolor{green!50!black}{-0.023$\pm$0.290} & \textcolor{green!50!black}{-0.019$\pm$0.244} \\
 & \textcolor{red}{2} & \textcolor{red}{0.002$\pm$0.126} & \textcolor{red}{0.008$\pm$0.154} & \textcolor{red}{0.014$\pm$0.209} & \textcolor{red}{0.010$\pm$0.194} \\
\hline
$\theta$ & \textcolor{blue}{0} & \textcolor{blue}{-0.087$\pm$2.918} & \textcolor{blue}{0.221$\pm$3.116} & \textcolor{blue}{0.195$\pm$3.422} & \textcolor{blue}{-11.021$\pm$29.242} \\
$[{\rm{degree}}]$ & \textcolor{green!50!black}{1} & \textcolor{green!50!black}{-0.445$\pm$2.777} & \textcolor{green!50!black}{-0.108$\pm$2.735} & \textcolor{green!50!black}{0.182$\pm$3.448} & \textcolor{green!50!black}{-13.564$\pm$29.765} \\
 & \textcolor{red}{2} & \textcolor{red}{-0.161$\pm$2.832} & \textcolor{red}{-0.183$\pm$4.619} & \textcolor{red}{0.359$\pm$4.882} & \textcolor{red}{-10.368$\pm$33.568} \\
\hline

\end{tabular}
}

\caption{Systemic offset and normalized median absolute deviation of recovered vs true parameters in SE3D fitting on toy model galaxies. The last four columns indicate observations of different (spatially resolved) wavelength coverage.
}
\label{tab:SE3D_Toy}
\end{table}

\subsection{Recovery on mock-observed toy models}
\label{sec:recov_toy}

\begin{figure*}
    \centering
    \includegraphics[width=\linewidth]{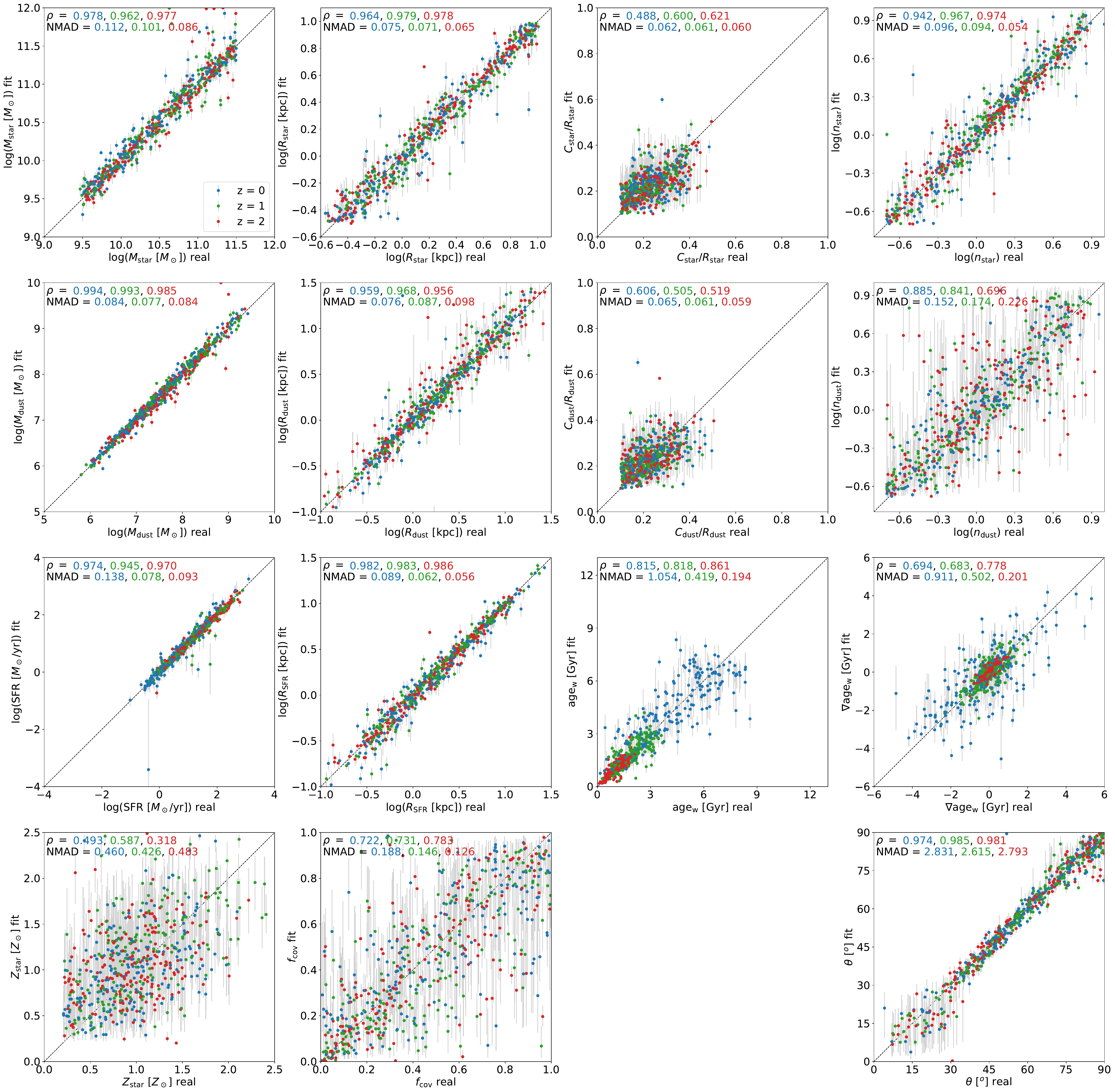}
    \caption{Recovered parameters from {\tt SE3D} fitting contrasted with true intrinsic toy model parameters for the filter band combination `IRres' and 0.05 dex uncertainty. We fit a sample of 600 toy model galaxies equally divided between three redshifts: z=0 (blue), z=1 (green), and z=2 (red).}
    \label{fig:SE3Dfit_varying_redshift}
\end{figure*}

\subsubsection{Fitting panchromatic and spatially resolved data}
\label{sec:recov_toy_default}

In first instance, we fit the mock-observed spectral distributions of the toy models described in Section\ \ref{sec:toymodels} using the full panchromatic and spatially resolved information as outlined in the top row (`IRres') of Table\ \ref{tab:filters}.  That is, the SED is well sampled from UV to sub-mm wavelengths, and spatially resolved diagnostics are assumed to be available in all but the {\it Spitzer} and {\it Herschel} bands.  To mimic observational measurement uncertainties, we perturb all fluxes, sizes, S\'{e}rsic indices and projected axis ratios by adding a Gaussian random variable in log-space, with $\sigma_{\rm perturb} = 0.05\ {\rm dex}$.  The same 0.05 dex is also adopted as measurement error on all data points in the fitting.  Variations in data quality and availability (e.g., wavelength coverage) are considered in the subsequent sections.

Figure\ \ref{fig:SE3Dfit_varying_redshift} contrasts best-fit values against ground truth for an array of relevant physical quantities.  In each panel, we mark the $z=0/1/2$ objects by different colours, and quote summary statistics on the strength of correlation between recovered and true properties (Spearman rank correlation coefficient, $\rho$) and their deviation (the normalized median absolution deviation, ${\rm NMAD} \equiv 1.4826 \times {\rm median}(|x_{\rm fit} - x_{\rm true}|)$).  In Table\ \ref{tab:SE3D_Toy} we break the latter metric down into the median systematic offset and scatter around that median offset.

We consider recovery performance for stellar mass and associated structural properties (top row), dust mass and associated structural properties (2$^{\rm nd}$ row), (resolved) stellar population related quantities (3$^{\rm rd}$ row), and finally stellar metallicity, birth cloud covering fraction and observer's viewing angles (bottom row).  We find the overall recovery accuracy to be good, with minimal systematic uncertainties and only modest scatter in all but a few parameters.  We remind the reader that this result is non-trivial, despite the test objects satisfying the same toy model description as also employed in our fitting.  First, the mock observations are generated with SKIRT, whereas the fitting is done using an emulator trained to mimic (albeit imperfectly) SKIRT functionality.  Secondly, the fitting is applied to a finite sampling of the intrinsic spectral distributions which moreover were perturbed by measurement errors.  In light of these considerations, the outcome of the recovery test is encouraging.  In particular, absolute quantities such as the total stellar mass, dust mass and SFR are recovered well, with a scatter of around $\sim 0.1$ dex.  We note that the SFR recovery actually improves with increasing redshift, presumably because shorter rest-wavelengths are probed, sensitive to unobscured young stellar populations.  Half-mass radii of stars and dust, the half-SFR radius and stellar S\'{e}rsic index are all similarly well recovered, to 0.1 dex or better, whereas a somewhat larger scatter is observed for the S\'{e}rsic index of the dust component.\footnote{We recognize that the flat measurement uncertainty of 0.05 dex, applied to all fluxes and global structural parameters regardless of instrument/waveband, may be somewhat unrealistic.  This approach was adopted primarily for simplicity of the argument.  In practice, the photometric accuracy will often be higher, whereas the ability to constrain a S\'{e}rsic index based on marginally resolved {\it JWST}/MIRI and ALMA observations may be more limited \citep[see, e.g.,][]{Tadaki2020,Magnelli2023}.}

Specific stellar population properties such as the mass-weighted age and difference between mass-weighted age outside vs inside $R_{\rm star}$ are more difficult to constrain, but nevertheless show a highly significant correlation between recovered and true properties, with small systematic offsets.  The strength of correlation drops when turning to stellar metallicity and birth cloud covering fraction, albeit formally remaining of statistical significance ($p \ll 0.01$).  The larger scatter is also reflected in broader posterior distributions (light gray error bars marking the central 68th percentile).  Conversely, we verified that fixing $Z_{\rm star}$ and/or $f_{\rm cov}$ to their true values while fitting does not significantly improve the recovery accuracy of other physical quantities. R26's Appendix C also illustrates the minimal impact stellar metallicity has on the predicted spectral distributions. Their sensitivity analysis suggests the $V-J$ colour may be the most sensitive observational probe, albeit trumped by other physical parameters impacting this diagnostic.  These findings confirm common wisdom in the literature that spectroscopy is required to confidently pin down stellar metallicities \citep[e.g.,][]{Gallazzi2005, Zahid2017}.

Finally, we point out that viewing angle ($\theta$) and disk thickness ($C/R$) could reasonably be expected to suffer some degeneracy.  Nevertheless, $\theta$ is recovered very well, with a scatter of less than $3^\circ$, and both $C_{\rm star}/R_{\rm star}$ and $C_{\rm dust}/R_{\rm dust}$ do not simply return the prior.  Instead, a modest yet significant correlation of $\rho \sim 0.5$ ($p \ll 0.01$) is observed for each.  We attribute this to the fact that disk thickness as well as viewing angle leave an imprint on more observables than merely the projected axis ratio (see the sensitivity analysis in R26).

\subsubsection{Varying data quality}
\label{sec:data_qual}

We repeated the recovery experiment twice using smaller measurement uncertainties, of 0.02 dex and 0.01 dex, respectively.  These variations did not yield noticeably reduced deviations between recovered and true parameter values.  Whereas this may appear surprising at first glance, there are two potential culprits for the lack of improvement: the inference may be limited by the finite wavelength sampling and/or the {\tt SE3D} emulator not perfectly reproducing SKIRT ground truth. Captured in the form of NMADs, the emulator accuracy for flux, size, S\'{e}rsic index and projected axis ratio SDs is 0.043, 0.030, 0.056 and 0.004 dex, respectively. For a more detailed quantification and assessment of emulator performance, we refer the reader to R26.  We tested the latter hypothesis by rerunning the experiment with mock observations now produced using the emulator as opposed to actual SKIRT RT calculations.  In this case, recovery accuracy does steadily improve when reducing mock errors from 0.05 to 0.01 dex.  We conclude that finite emulator accuracy does introduce a form of template mismatch which acts as a floor to the precision with which intrinsic parameters can be recovered from SKIRT mock observations.

\subsubsection{Varying (spatially resolved) wavelength coverage}
\label{sec:recov_toy_wavelength}

Besides our default data coverage (dubbed `IRres'), we experimented with three alternative realizations of the input dataset, detailed in Table\ \ref{tab:filters}:
\begin{itemize}
\item[\textbullet] IRint: no resolved diagnostics longwards of F444W
\item[\textbullet] noIR: lack of any observational constraints longwards of F444W
\item[\textbullet] nores: lack of any resolved diagnostics
\end{itemize}

The recovery statistics for each of these settings are listed in Table\ \ref{tab:SE3D_Toy}.  Omitting resolved information from tracers of dust re-emission ('IRint') has relatively little impact on any of the stellar properties.  It does increase the scatter in recovered versus true total dust mass, by a factor $\sim 1.2 - 1.5$ depending on redshift.  Unsurprisingly, the impact on resolved dust properties is larger, with a factor $\sim 1.3 - 2.0$ boost in deviations between recovered and true dust structural parameters ($R_{\rm dust}$, $C_{\rm dust}/R_{\rm dust}$, $n_{\rm dust}$).  That said, systematic offsets remain low, and $R_{\rm dust}$ is still recovered to within 0.2 dex, despite the lack of direct probes of resolved dust in emission.  We attribute this to the fact that the dust geometry also impacts the resolved properties at wavelengths where stellar emission dominates (see, e.g., \citealt{Zhang2023}, R26), to which the fitting procedure had access.

Absence of IR constraints altogether (`noIR') further increases the scatter in most quantities, both those related to dust and to stars (see Table\ \ref{tab:SE3D_Toy}).  While systematic offsets generally remain low (below 0.06 dex), they do on occasion take higher absolute values than in the `IRres' case.

Finally, we consider what can be inferred about the physical make-up of our test objects if no resolved information were available altogether (`nores').  This equates to using the same input information as available to standard SED modelling, i.e., omitting any Spectral Structural Distribution input.  However, in executing this test we still leave the fitting procedure with the same freedom as before, thus including the ability to adjust structural parameters freely in an attempt to find the best-fitting SED.  Compared to the `IRres' case, systematic offsets and scatter increase by typical factors of $\sim 2.0$ and $\sim 1.5$, respectively, for bulk quantities such as $M_{\rm star}$, $M_{\rm dust}$ and SFR.  For quantities that convey resolved properties, the systematic and random uncertainties increase by a typical factor of $\sim 2.5$ compared to `IRres'.  That said, they still show a remarkably consistent recovery of stellar, dust and SFR radii (see Figure\ \ref{fig:SE3Dfit_nores} in Appendix\ \ref{app:nores}).  Systematic offsets for these parameters remain below 0.1 dex, and the associated scatter remains $\lesssim 0.2$ dex for $R_{\rm star}$ and $R_{\rm SFR}$, and $\lesssim 0.35$ dex for $R_{\rm dust}$. We emphasize that this is the case despite a lack of built-in correlations among the toy model galaxies or within the SKIRT library on which our emulator was trained between on the one hand galaxy size and on the other hand parameters defining the star formation history, or the total stellar or dust mass.

A clue regarding the reason for the persistent recovery of toy model sizes despite the complete absence of spatially resolved information arises from a test conducted using only SED information as input to the fitting, but now restricted to photometry up to F444W (i.e., no sampling of dust in emission).  From this, we observe the recovery of structural parameters to drastically break down.  

We conclude that {\it panchromatic} SEDs retain information on the spatial extent of galaxy components, for at least two reasons: First, for a given dust mass, which is constrained most tightly by the Rayleigh-Jeans tail of the far-IR SED, the galaxy size will determine the dust column and will thus impact how effectively this dust attenuates and reddens the short-wavelength photometry.  Secondly, more compact configurations result in higher dust temperatures, modulating the far-IR SED shape.    Perhaps counterintuitively, constraints on galaxy size are thus achievable based on panchromatic SED information alone, at least in the context of the family of toy model galaxies considered in this exercise.  The latter is an important caveat, as template mismatch between real galaxies and the toy model description employed in {\tt SE3D} fitting may render the translation from SED input to structural information more challenging in the absence resolved observations.  The ability to retrieve structural information without resolved observations should also not be overstated, as Figure\ \ref{fig:SE3Dfit_nores} for example clearly depicts that the recovery of S\'{e}rsic indices and viewing angle become effectively unconstrained, and that the returned thickness of stellar and dust disks simply echoes the prior.

\subsection{Recovery on mock-observed TNG galaxies}
\label{sec:recov_TNG}

\begin{table}
\centering 
\resizebox{\linewidth}{!}{
\begin{tabular}{c|c|c|c|c|c}
\hline \hline

TNG & Redshift & IRres & IRint & noIR & nores \\
\hline

$\log(M_{\rm{star}})$ & \textcolor{blue}{0} & \textcolor{blue}{-0.045$\pm$0.077} & \textcolor{blue}{-0.043$\pm$0.076} & \textcolor{blue}{-0.047$\pm$0.097} & \textcolor{blue}{-0.018$\pm$0.076} \\
$[\log(M_\odot)]$ & \textcolor{green!50!black}{1} & \textcolor{green!50!black}{0.027$\pm$0.101} & \textcolor{green!50!black}{0.039$\pm$0.097} & \textcolor{green!50!black}{0.037$\pm$0.137} & \textcolor{green!50!black}{-0.005$\pm$0.114} \\
 & \textcolor{red}{2} & \textcolor{red}{-0.018$\pm$0.135} & \textcolor{red}{-0.015$\pm$0.105} & \textcolor{red}{-0.129$\pm$0.157} & \textcolor{red}{-0.044$\pm$0.102} \\
\hline
$\log(R_{\rm{star}})$ & \textcolor{blue}{0} & \textcolor{blue}{-0.026$\pm$0.078} & \textcolor{blue}{-0.030$\pm$0.084} & \textcolor{blue}{-0.048$\pm$0.097} & \textcolor{blue}{0.014$\pm$0.176} \\
$[\log({\rm{kpc}})]$ & \textcolor{green!50!black}{1} & \textcolor{green!50!black}{0.031$\pm$0.120} & \textcolor{green!50!black}{0.017$\pm$0.103} & \textcolor{green!50!black}{0.020$\pm$0.159} & \textcolor{green!50!black}{0.069$\pm$0.202} \\
 & \textcolor{red}{2} & \textcolor{red}{0.105$\pm$0.143} & \textcolor{red}{0.035$\pm$0.143} & \textcolor{red}{0.231$\pm$0.230} & \textcolor{red}{0.107$\pm$0.175} \\
\hline
$C_{\rm{star}}/R_{\rm{star}}$ & \textcolor{blue}{0} & \textcolor{blue}{0.005$\pm$0.064} & \textcolor{blue}{0.006$\pm$0.085} & \textcolor{blue}{0.009$\pm$0.091} & \textcolor{blue}{0.043$\pm$0.103} \\
 & \textcolor{green!50!black}{1} & \textcolor{green!50!black}{-0.016$\pm$0.092} & \textcolor{green!50!black}{-0.011$\pm$0.091} & \textcolor{green!50!black}{-0.013$\pm$0.092} & \textcolor{green!50!black}{0.024$\pm$0.089} \\
 & \textcolor{red}{2} & \textcolor{red}{-0.009$\pm$0.079} & \textcolor{red}{-0.011$\pm$0.094} & \textcolor{red}{-0.018$\pm$0.092} & \textcolor{red}{0.013$\pm$0.071} \\
\hline
$\log(n_{\rm{star}})$ & \textcolor{blue}{0} & \textcolor{blue}{-0.117$\pm$0.184} & \textcolor{blue}{-0.094$\pm$0.200} & \textcolor{blue}{-0.064$\pm$0.224} & \textcolor{blue}{-0.462$\pm$0.308} \\
$[{\rm{dex}}]$ & \textcolor{green!50!black}{1} & \textcolor{green!50!black}{-0.094$\pm$0.270} & \textcolor{green!50!black}{-0.058$\pm$0.237} & \textcolor{green!50!black}{-0.018$\pm$0.299} & \textcolor{green!50!black}{-0.248$\pm$0.370} \\
 & \textcolor{red}{2} & \textcolor{red}{0.025$\pm$0.287} & \textcolor{red}{0.103$\pm$0.234} & \textcolor{red}{-0.056$\pm$0.316} & \textcolor{red}{0.052$\pm$0.278} \\
\hline
$\log(M_{\rm{dust}})$ & \textcolor{blue}{0} & \textcolor{blue}{0.089$\pm$0.062} & \textcolor{blue}{0.100$\pm$0.057} & \textcolor{blue}{0.067$\pm$0.237} & \textcolor{blue}{0.115$\pm$0.060} \\
$[\log(M_\odot)]$ & \textcolor{green!50!black}{1} & \textcolor{green!50!black}{0.042$\pm$0.050} & \textcolor{green!50!black}{0.038$\pm$0.066} & \textcolor{green!50!black}{-0.112$\pm$0.252} & \textcolor{green!50!black}{0.087$\pm$0.069} \\
 & \textcolor{red}{2} & \textcolor{red}{0.044$\pm$0.051} & \textcolor{red}{0.011$\pm$0.065} & \textcolor{red}{-0.195$\pm$0.300} & \textcolor{red}{0.055$\pm$0.056} \\
\hline
$\log(R_{\rm{dust}})$ & \textcolor{blue}{0} & \textcolor{blue}{0.014$\pm$0.061} & \textcolor{blue}{0.012$\pm$0.099} & \textcolor{blue}{0.008$\pm$0.159} & \textcolor{blue}{0.002$\pm$0.135} \\
$[\log({\rm{kpc}})]$ & \textcolor{green!50!black}{1} & \textcolor{green!50!black}{0.037$\pm$0.081} & \textcolor{green!50!black}{0.002$\pm$0.165} & \textcolor{green!50!black}{-0.076$\pm$0.228} & \textcolor{green!50!black}{0.036$\pm$0.253} \\
 & \textcolor{red}{2} & \textcolor{red}{0.028$\pm$0.114} & \textcolor{red}{-0.085$\pm$0.210} & \textcolor{red}{0.135$\pm$0.408} & \textcolor{red}{-0.167$\pm$0.309} \\
\hline
$C_{\rm{dust}}/R_{\rm{dust}}$ & \textcolor{blue}{0} & \textcolor{blue}{0.066$\pm$0.072} & \textcolor{blue}{0.057$\pm$0.067} & \textcolor{blue}{0.060$\pm$0.074} & \textcolor{blue}{0.111$\pm$0.059} \\
 & \textcolor{green!50!black}{1} & \textcolor{green!50!black}{0.024$\pm$0.098} & \textcolor{green!50!black}{0.053$\pm$0.082} & \textcolor{green!50!black}{0.061$\pm$0.079} & \textcolor{green!50!black}{0.060$\pm$0.086} \\
 & \textcolor{red}{2} & \textcolor{red}{-0.017$\pm$0.074} & \textcolor{red}{0.044$\pm$0.106} & \textcolor{red}{0.060$\pm$0.095} & \textcolor{red}{0.035$\pm$0.093} \\
\hline
$\log(n_{\rm{dust}})$ & \textcolor{blue}{0} & \textcolor{blue}{0.110$\pm$0.143} & \textcolor{blue}{0.213$\pm$0.311} & \textcolor{blue}{0.285$\pm$0.338} & \textcolor{blue}{0.621$\pm$0.350} \\
$[{\rm{dex}}]$ & \textcolor{green!50!black}{1} & \textcolor{green!50!black}{0.039$\pm$0.207} & \textcolor{green!50!black}{0.193$\pm$0.644} & \textcolor{green!50!black}{0.266$\pm$0.595} & \textcolor{green!50!black}{0.722$\pm$0.371} \\
 & \textcolor{red}{2} & \textcolor{red}{0.276$\pm$0.402} & \textcolor{red}{0.700$\pm$0.569} & \textcolor{red}{0.278$\pm$0.553} & \textcolor{red}{0.591$\pm$0.426} \\
\hline
$\log({\rm{SFR}})$ & \textcolor{blue}{0} & \textcolor{blue}{0.072$\pm$0.173} & \textcolor{blue}{0.078$\pm$0.166} & \textcolor{blue}{0.116$\pm$0.182} & \textcolor{blue}{-0.028$\pm$0.071} \\
$[\log(M_\odot/{\rm{yr}})]$ & \textcolor{green!50!black}{1} & \textcolor{green!50!black}{-0.020$\pm$0.099} & \textcolor{green!50!black}{-0.018$\pm$0.073} & \textcolor{green!50!black}{-0.090$\pm$0.171} & \textcolor{green!50!black}{-0.010$\pm$0.076} \\
 & \textcolor{red}{2} & \textcolor{red}{-0.007$\pm$0.095} & \textcolor{red}{0.018$\pm$0.076} & \textcolor{red}{-0.251$\pm$0.253} & \textcolor{red}{0.041$\pm$0.071} \\
\hline
$\log(R_{\rm{SFR}})$ & \textcolor{blue}{0} & \textcolor{blue}{-0.008$\pm$0.082} & \textcolor{blue}{-0.010$\pm$0.091} & \textcolor{blue}{-0.010$\pm$0.096} & \textcolor{blue}{-0.237$\pm$0.153} \\
$[\log({\rm{kpc}})]$ & \textcolor{green!50!black}{1} & \textcolor{green!50!black}{-0.040$\pm$0.097} & \textcolor{green!50!black}{-0.110$\pm$0.136} & \textcolor{green!50!black}{-0.053$\pm$0.174} & \textcolor{green!50!black}{-0.189$\pm$0.150} \\
 & \textcolor{red}{2} & \textcolor{red}{-0.024$\pm$0.101} & \textcolor{red}{-0.154$\pm$0.197} & \textcolor{red}{0.134$\pm$0.239} & \textcolor{red}{-0.185$\pm$0.205} \\
\hline
${\rm{age}}_w$ & \textcolor{blue}{0} & \textcolor{blue}{-0.715$\pm$1.025} & \textcolor{blue}{-0.679$\pm$1.032} & \textcolor{blue}{-0.708$\pm$1.061} & \textcolor{blue}{-0.115$\pm$1.021} \\
$[{\rm{Gyr}}]$ & \textcolor{green!50!black}{1} & \textcolor{green!50!black}{0.100$\pm$0.534} & \textcolor{green!50!black}{0.126$\pm$0.544} & \textcolor{green!50!black}{0.300$\pm$0.601} & \textcolor{green!50!black}{0.082$\pm$0.495} \\
 & \textcolor{red}{2} & \textcolor{red}{-0.001$\pm$0.237} & \textcolor{red}{-0.017$\pm$0.190} & \textcolor{red}{0.190$\pm$0.383} & \textcolor{red}{0.008$\pm$0.152} \\
\hline
$\nabla {\rm{Age}}_{\rm{w}}$ & \textcolor{blue}{0} & \textcolor{blue}{-0.408$\pm$1.697} & \textcolor{blue}{-0.192$\pm$1.773} & \textcolor{blue}{-0.363$\pm$1.703} & \textcolor{blue}{1.802$\pm$1.125} \\
$[{\rm{Gyr}}]$ & \textcolor{green!50!black}{1} & \textcolor{green!50!black}{0.086$\pm$0.881} & \textcolor{green!50!black}{0.245$\pm$0.739} & \textcolor{green!50!black}{0.214$\pm$0.698} & \textcolor{green!50!black}{0.601$\pm$0.514} \\
 & \textcolor{red}{2} & \textcolor{red}{0.136$\pm$0.422} & \textcolor{red}{0.216$\pm$0.273} & \textcolor{red}{-0.003$\pm$0.335} & \textcolor{red}{0.392$\pm$0.321} \\
\hline
$Z$ & \textcolor{blue}{0} & \textcolor{blue}{0.104$\pm$0.229} & \textcolor{blue}{0.141$\pm$0.263} & \textcolor{blue}{0.271$\pm$0.323} & \textcolor{blue}{0.100$\pm$0.238} \\
$[Z_\odot]$ & \textcolor{green!50!black}{1} & \textcolor{green!50!black}{-0.190$\pm$0.381} & \textcolor{green!50!black}{-0.212$\pm$0.379} & \textcolor{green!50!black}{-0.026$\pm$0.441} & \textcolor{green!50!black}{-0.140$\pm$0.323} \\
 & \textcolor{red}{2} & \textcolor{red}{-0.161$\pm$0.338} & \textcolor{red}{-0.166$\pm$0.312} & \textcolor{red}{-0.129$\pm$0.321} & \textcolor{red}{-0.187$\pm$0.234} \\
\hline
$\theta$ & \textcolor{blue}{0} & \textcolor{blue}{0.298$\pm$2.331} & \textcolor{blue}{0.086$\pm$2.673} & \textcolor{blue}{0.306$\pm$2.842} & \textcolor{blue}{-19.617$\pm$26.708} \\
$[{\rm{degree}}]$ & \textcolor{green!50!black}{1} & \textcolor{green!50!black}{-1.408$\pm$3.665} & \textcolor{green!50!black}{-0.907$\pm$3.652} & \textcolor{green!50!black}{-0.646$\pm$3.259} & \textcolor{green!50!black}{-16.676$\pm$28.952} \\
 & \textcolor{red}{2} & \textcolor{red}{-1.147$\pm$3.313} & \textcolor{red}{-0.640$\pm$4.167} & \textcolor{red}{-0.948$\pm$3.352} & \textcolor{red}{-18.893$\pm$28.990} \\
\hline

\end{tabular}
}

\caption{Systemic offset and normalized median absolute deviation of recovered versus true parameters in SE3D fitting on TNG galaxies. The last four columns indicate observations of different (spatially resolved) wavelength coverage.
}
\label{tab:SE3D_TNG}
\end{table}

\begin{figure*}
    \centering
    \includegraphics[width=\linewidth]{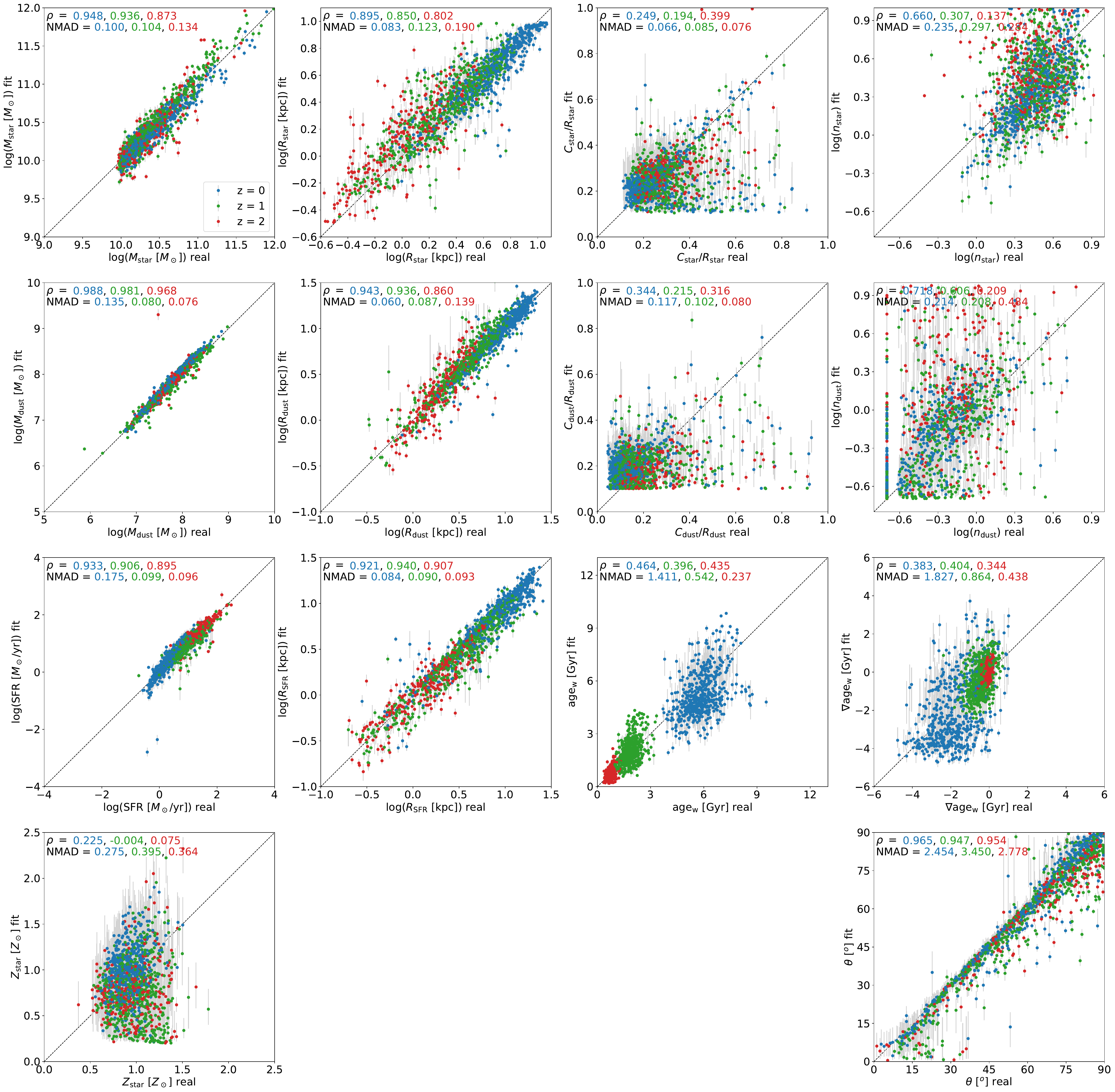}
    \caption{{\tt SE3D} fitting on TNG galaxies with filter band combination `IRres' and uncertainty 0.05 dex. TNG galaxies at redshift $z$ = 0, 1, 2 are presented in blue, green and red colours, respectively.}
    \label{fig:SE3Dfit_TNG}
\end{figure*}

We now turn to an equivalent recovery analysis, but applied to mock-observed TNG rather than toy model galaxies.  We contrast best-fit to intrinsic (i.e., `true') parameters in Figure\ \ref{fig:SE3Dfit_TNG} for our fiducial wavelength sampling, and for all filter combinations specified in Table\ \ref{tab:filters} we record statistics on the recovery accuracy in Table\ \ref{tab:SE3D_TNG}.

\subsubsection{Recovery results for TNG}

The total integrated stellar mass, dust mass and star formation rates of TNG galaxies are recovered well, with systematics well below 0.1 dex and comparable scatter as found for the recovery on toy model galaxies ($\sim 0.1$ dex).

The effective radii of the stellar mass, dust mass and SFR distribution are likewise in good agreement with the intrinsic sizes, with systematics of 0.04 dex or less for all but $R_{\rm star}$ at $z=2$ (for which we find +0.09 dex). The scatter in recovered versus intrinsic sizes ranges between 0.06 and 0.15 dex, depending on redshift and component (stars/dust/SFR).

For the recovery of $n_{\rm star}$ and $n_{\rm dust}$, we retrieve a comparatively larger scatter than seen in the toy model case, of order $\sim 0.2 - 0.4$ dex, with also more substantial systematic offsets, ranging from -0.11 to +0.27 dex, depending on which redshift and component (stars/dust) is considered.  Here, we remind the reader that `truth' in this case corresponds to the S\'{e}rsic index that most closely approximates the TNG galaxy's actual mass distribution in the {\it xy} plane (see Section\ \ref{sec:true}).  The actual TNG profiles are not per se well described by a S\'{e}rsic functional form.  We also note that the S\'{e}rsic indices that best describe the SFGs in our TNG sample follow a distinct distribution from those adopted when composing our toy model test objects.  For our toy models, we drew the S\'{e}rsic indices of stellar and dust components from Gaussian distributions in $\log(n)$ with a scatter of 0.6 dex, centred on the exponential disk case ($n = 1$) for dust and only slightly cuspier ($n = 1.4$) for the stars (see Table 1 in R26).  The median indices measured for massive SFGs in TNG50 are $\langle n_{\rm star} \rangle = 2.9$ (typically considered bulge-dominated in photometric decomposition studies) and $\langle n_{\rm dust} \rangle = 0.5$ (closer to a Gaussian radial profile).  For a non-negligible fraction of TNG galaxies at $z = 0$ and $z = 1$, the radial dust profile was shallow enough (or sometimes featured a central hole) such that the determination of $n_{\rm dust, real}$ hit the lower bound of $n = 0.2$ that we imposed. 

Considering the mass-weighted age and its radial gradient, we find that the overall trend of increasing ${\rm age_w}$ and more negative $\nabla {\rm age_w}$ towards later times is recovered.  At any given epoch, however, the correlation between best-fit and intrinsic age (gradient), while formally significant, is weak ($\rho_{\rm Spearman} \sim 0.4$).  We investigate the physical origin of their relatively poorer recovery more closely in Section\ \ref{sec:discussion}.  At face value, we conclude that measures of mass-weighted age and age gradients on an individual object basis should be treated with caution, but in aggregate they can still be of use to study population-based redshift trends such as those discussed in Section\ \ref{sec:intrinsic}. 

Finally, regarding disk thickness and viewing angle, we find the inclinations are recovered well, to within a few degrees, despite the intrinsic thickness of stellar and dust distributions not being fixed to their true values.  Instead, they were left to be fitted freely, with only guidance via a fiducial Gaussian prior of $(\mu,\ \sigma) = (0.2,\ 0.1)$, which straddles the actual thickness distributions of the stellar and dust components as presented in Figure\ \ref{fig:TNG_params_dist}.\footnote{Observationally, such a prior on the stellar thickness could be informed by an analysis of the projected axis ratio distribution of an ensemble of galaxies \citep[see, e.g.,][]{van2014b}.  It is more challenging, however, to derive equivalent constraints independently for the dust component, as large number statistics of spatially resolved dust continuum maps would be required.}  In this context, it is interesting to note that, while to first order best-fitting values of $C_{\rm star}/R_{\rm star}$ and $C_{\rm dust}/R_{\rm dust}$ do not veer away much from the prior, the recovery analysis does correctly infer two aspects regarding the geometric thickness.  First, recovered $C_{\rm dust}/R_{\rm dust}$ are on average smaller than $C_{\rm star}/R_{\rm star}$, pointing to a thinner ISM than stellar geometry.  Secondly, the typical best-fit $C_{\rm star}/R_{\rm star}$ increases with increasing redshift, a trend which is also present intrinsically and was highlighted in Figure\ \ref{fig:TNG_params_dist}.

We conclude that, despite the more complex SFHs and spatial distributions of TNG galaxies, {\tt SE3D} performs encouragingly well in recovering several of the key intrinsic physical characteristics of the test objects studied.

\subsubsection{Covariance between derived parameters}
\label{sec:covariance}
In any multi-dimensional modelling of data, the results may be subject to covariances between inferred properties.  We investigated the presence of degeneracies between derived parameters in two ways: First, we considered for each test object and each pair of inferred properties to which degree walker positions stored by the {\tt emcee} algorithm (after discarding a burn-in phase) were correlated, as opposed to displaying a random scatter plot.  These metrics were subsequently averaged over the sample of test objects.  As a second method of assessment, we quantified the Spearman's rank correlation coefficient between the residuals from truth for one parameter and those for another parameter, again evaluating all possible parameter pairs.  In both cases, we ranked the parameter pairs in order of descending amplitude of the correlation coefficient, and found a consistent set of a handful of parameter pairings to exhibit significant covariance (correlation coefficients of $\rho \gtrsim 0.6$).  These include for example the stellar mass and mass-weighted age (older populations requiring more mass to reproduce the same observed brightness), stellar thickness and inclination angle (thicker structures requiring a higher inclination to reproduce the same observed axis ratio), and stellar size and age gradient (a more compact stellar distribution requiring younger outskirts to reproduce the same observed half-light radius).  Similarly, stellar size and stellar S\'{e}rsic index ranks among the more covariant parameter pairings (a more compact stellar distribution can to some extent be compensated by a higher S\'{e}rsic index to yield a similar match to the observed half-light radii and S\'{e}rsic light profiles).  A similar exercise conducted on the set of toy model test objects covered in Section\ \ref{sec:recov_toy} yields a consistent set of parameter pairings that show more signs of covariance.  Of note is that, while the correlation for these pairings is relatively strong, the degeneracy only extends over a limited dynamic range compared to the range of galaxy properties considered in our recovery analysis, thus leading to the modest confidence intervals displayed in Figures\ \ref{fig:SE3Dfit_varying_redshift} and\ \ref{fig:SE3Dfit_TNG}, which were extracted from the marginalised posteriors.  We also note that in experiments with reduced input data (Section\ \ref{sec:recov_toy_wavelength}) the derived levels of covariance did not increase, but if anything reduced.

\section{Discussion}
\label{sec:discussion}

In Section\ \ref{sec:results}, we evaluated the radial age gradients of massive SFGs since cosmic noon in TNG50, and inspected the relative shape (thickness) and radial extent of their stellar and dust components.  We also presented their observable characteristics in $UVJ$ and IRX-$\beta$ space.

Unlike the RT results for our sample of TNG50 SFGs, the library of toy model galaxies we composed does span the full range in $UVJ$ and IRX-$\beta$ covered by observed galaxies.  We therefore proceeded to conduct a recovery analysis of intrinsic physical properties via {\tt SE3D} fitting on both test objects of toy model and simulated nature.  Generally speaking, with the exception of certain properties such as $M_{\rm star}$, $M_{\rm dust}$, $R_{\rm star}$ and $R_{\rm dust}$ which are well-recovered for both TNG and toy models, the recovery accuracy on TNG test objects is relatively poorer.

In the following, we elaborate briefly on the realism of mock-observed simulated galaxies (Section\ \ref{sec:obssim}), take a deeper dive into the consequences of model mismatch (Section\ \ref{sec:modelmismatch}) and finally consider potential fruitful extensions to {\tt SE3D} in its current incarnation (Section\ \ref{sec:extensions}).

\subsection{Realistic observables for simulated galaxies}
\label{sec:obssim}

The challenge to reconcile simulations with the full range of $UVJ$ colours exhibited by observed galaxies was first noted by \citet{Wuyts2009b}, and has not been fully resolved in later cosmological simulations with more sophisticated RT treatment either \citep{Donnari2019,Akins2022,Gebek2025}.  If anything, more precise photometry and photometric redshifts for the most dust reddened SFGs enabled by {\it JWST} has amplified this tension \citep{vanderWel2025}.  Where successes were achieved in producing a larger spread of $UVJ$ colours, the synthetic colours did not exhibit the same dependence on galaxy stellar mass as observed \citep{Akins2022}.  To some degree, the challenge to achieve sufficient reddening (and efficient reddening, not just obscuration), is also reflected in the IRX-$\beta$ diagram (see Section\ \ref{sec:obsdiagrams} and, e.g., \citealt{Schulz2020}).

While we do not offer a solution to this conundrum, we point out that a priori the issue could be attributed to one (or multiple) of the following factors being off from the real conditions: (1) the amount of dust assigned to the simulated galaxies; (2) the global spatial distribution of dust with respect to that of the stars; (3) the local dust distribution on subgalactic and birth cloud scales; (4) the intrinsic emission (and spatial distribution) of the stellar populations. For a more elaborate display of the impact on SDs from varying global or local parameters, we refer the reader to Appendix C in R26.

The last factor is likely of lesser concern: TNG galaxies have been shown to reasonably reproduce the evolving galaxy stellar mass function, star-forming main sequence and even resolved star formation profiles \citep{Pillepich2018b,Nelson2021}.  For a given distribution of stellar populations, different stellar population synthesis models induce colour changes that, while non-negligible, are overall modest \citep[e.g.,][]{Akins2022}. 

The amount of dust and its global distribution are inherently tied to the method used to `paint' dust onto gas particles, making use of the mass and metal distribution of cold/dense gas particles as traced in the simulation.  In Appendix\ \ref{app:dustassignment}, we provide a brief overview of different methods adopted in the literature and consider their impact on the intrinsic dust properties, as well as the observable $UVJ$ and IRX-$\beta$ diagnostics.  For the variations explored, changes to the observables again are non-negligible but insufficient to account for the reddest $V-J$ colours of observed galaxies.  We stress that these experiments only vary the method used to assign dust; they take the input gas and metal content from the simulation as a given.  Compared to empirically established gas scaling relations \citep{Tacconi2020}, cosmological simulations such as TNG tend to feature a shallower evolution in galaxies' cold gas fraction.  Combined with a reduced gas-phase metallicity at earlier epochs, this then yields a striking lack of evolution in the inferred dust-to-stellar mass ratios compared to observations (see also \citealt{Millard2021}).  The typical stellar mass of SFGs in our $z=0$ and $z=2$ TNG samples is $\log(M_{\rm star}) = 10.3$.  At this mass, the increase in $M_{\rm dust}/M_{\rm star}$ from $z=0$ to $z=2$ as seen in observations amounts to $0.9 \pm 0.2$ dex, depending on the specific galaxy samples and/or scaling relations considered \citep{Casasola2020, DeLooze2020, Donevski2020, Tacconi2020}.  In contrast, using our fiducial dust assignment method, we derive for similar-mass SFGs in TNG a typical $\log(M_{\rm dust}/M_{\rm star}) \sim -2.6$ with less than 0.1 dex redshift evolution and a stable central 68th percentile range of $[-3.0, -2.4]$, as such sitting in between the empirical estimates at $z = 0$ (-3.1) and $z=2$ (-2.1).\footnote{We note that, in line with the observational scaling relations, the $M_{\rm dust}/M_{\rm star}$ inferred from TNG do generally decline towards the high-mass end, with substantial scatter at any mass.}  Appendix\ \ref{app:dustassignment} illustrates that alternative choices for dust assignment can yield more $M_{\rm dust}/M_{\rm star}$ evolution, by up to 0.3 dex, but this is the result of a reduced dust content at $z=0$, not an increased one at $z=2$.  

Difficulties in reconciling observed and simulated $UVJ$ and IRX-$\beta$ diagrams may thus relate to the globally less ISM-rich nature of galaxies produced in cosmological simulations, although an additional impact from global or local star-dust geometries cannot be excluded.  \citet{Gebek2025} experimented with artificially boosting the dust-to-stellar mass ratio to a sizeable $\log(M_{\rm dust}/M_{\rm star}) = -1.7$, and while obtaining high $A_V$ ($\sim 2$) found the $V-J$ reddening to still remain modest ($A_V - A_J \sim 0.4$). Ultimately, simulations which explicitly trace dust formation and destruction \citep[e.g.,][]{Schaye2026}, and which not only allow the amount of dust but also its spatial distribution to decouple from that of the gaseous metals \citep[e.g.,][]{Li2019} may offer new insights on the evolving dust content and star-dust geometries, and on the resulting colours.

Finally, using a set of 16 test objects from our TNG50 massive SFG sample, we evaluated the impact of changing the adopted birth cloud covering fraction from 50\% to 100\%.  To this end, we selected objects which represent a spread across the synthetic $UVJ$ diagram generated with our fiducial settings.  Changes in the resulting observables remain limited to $\sim 0.080$ in $U-V$, $\sim 0.013$ in $V-J$, $\sim 0.262$ in $\beta$ and $\sim 0.179$ in IRX.  More drastic changes to the birth cloud treatment (such as increasing their lifetimes by up to 2 orders of magnitude) may be needed to reproduce the colours of the most dust reddened cosmic noon SFGs \citep{Gebek2025}.

\subsection{Model mismatch}
\label{sec:modelmismatch}

Leaving the realism of TNG synthetic observables aside, we now consider the role of model mismatch (a.k.a. template mismatch) when fitting the mock-observed SDs of TNG galaxies with {\tt SE3D}.  Our approach effectively aims to determine which toy model description best reproduces the wavelength-dependent flux, size, S\'{e}rsic index and projected axis ratio of a given TNG galaxy.  In practice, however, our simplistic models cannot capture the full complexity of the simulated galaxies, an effect we refer to as model mismatch.  In order to more systematically address its impact, we distinguish between SFH mismatch (including its radial dependence), mismatch due to the assumed mono-metallicity stellar populations in our toy models, mismatch in the shape of the radial density profiles, and finally mismatch due to the azimuthally smooth nature of the diffuse stellar and dust components of our toy models.

We investigate the impact of model mismatch in two ways: first using the fitting results presented in Section\ \ref{sec:recov_TNG}, and secondly by artificially simplifying one feature of a set of TNG test objects at a time, and repeating the recovery experiment.

\subsubsection{Recovery accuracy as a function of model mismatch}
\label{sec:accuracy_vs_mismatch}

To quantitatively capture the overall quality with which physical properties of a given TNG galaxy are recovered, we introduce an overall `recovery residual' metric computed by summing in quadrature the normalized residuals for the different properties displayed in Figure\ \ref{fig:SE3Dfit_TNG}:
\begin{equation}
\mathcal{R} = \sqrt{\sum_i^{N_{\rm properties}} \frac{(X_{\rm fit, i} - X_{\rm real, i})^2}{N_{\rm properties}\Delta X_i^2}},
\end{equation}
where $\Delta X_i$ stands for the dynamic range of the given quantity and its inclusion in the denominator serves to assign residuals for different physical quantities (often of different dimensions also) approximately equal weight in the overall residual metric.  $\mathcal{R}$ is thus a dimensionless quantity with no concrete physical meaning, but it has the merit of being larger for fits that failed to properly recover the intrinsic parameters.

\begin{figure}
    \centering
    \includegraphics[width=\linewidth]{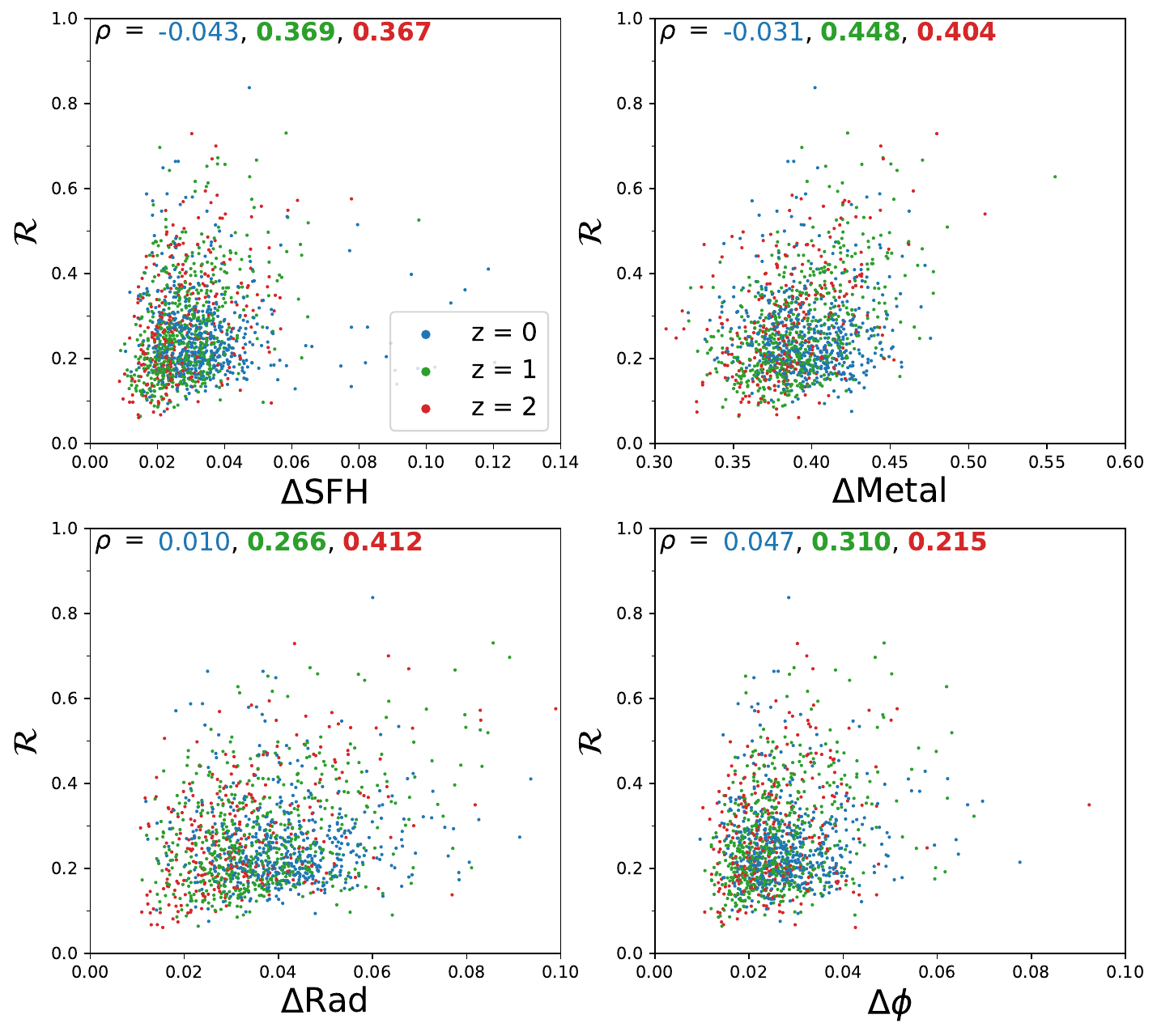}
    \caption{Recovery residuals $\mathcal{R}$ versus different aspects of ``model mismatch'': $\Delta {\rm SFH}$, $\Delta {\rm Metal}$, $\Delta {\rm Rad}$, and $\Delta \phi$. TNG galaxies at redshift $z$ = 0, 1, 2 are presented in blue, green and red colours, respectively. }
    \label{fig:residuals}
\end{figure}

Besides the `recovery residuals', we also quantify for each TNG galaxy in our sample the degree of model mismatch by defining an `SFH mismatch' ($\Delta {\rm SFH}$), `metallicity mismatch' ($\Delta {\rm Metal}$), `radial mismatch' ($\Delta {\rm Rad}$), and `azimuthal mismatch' ($\Delta \phi$).  We determine these quantities by comparing the particle level TNG properties to the closest toy model galaxy (see Section\ \ref{sec:true} for how this is identified).  The radial and azimuthal model mismatch can be quantified for both stars and dust separately, but for simplicity we combine them here into single parameters $\Delta {\rm Rad} = \sqrt{(\Delta {\rm Rad^2_{star}} + \Delta {\rm Rad^2_{dust}})/2}$ and $\Delta \phi = \sqrt{(\Delta \phi _{\rm star}^2 + \Delta \phi _{\rm dust}^2)/2}$.  We define $\Delta {\rm Metal}$ and $\Delta {\rm Rad}$ as the Cram\'{e}r-von Mises statistic $\omega$ \citep{Anderson1962}, effectively summing the residuals between TNG and toy model cumulative distribution functions (CDFs) in quadrature:
\begin{equation}
    \omega = \sqrt{ \sum {\left[ {\rm{ (CDF_{TNG}-CDF_{model})^2 \Delta CDF_{TNG} }} \right]}},
\label{eq:CvM}
\end{equation}
where ${\rm \Delta CDF_{TNG}}$ represents the discrete step in the TNG CDF at each respective point of the summation.  In the case of the metallicity mismatch, ${\rm CDF_{TNG}}$ refers to the CDF of stellar particle metallicities of the simulated TNG galaxy, whereas the CDF for our toy model is a step function (i.e., all stars were assigned identical metallicities).  In the case of the radial mismatch metric, ${\rm CDF_{TNG}}$ and ${\rm CDF_{model}}$ represent the cumulative radial mass profile of stars (or dust) for the TNG galaxy or best-fitting toy model, respectively. 

For $\Delta {\rm{SFH}}$, we similarly use the Cram\'{e}r-von Mises statistic $\omega$ to quantify how well the closest toy model reproduces a TNG galaxy, but now do so by effectively summing up the residuals of a 2D CDF in the age -- radius plane in quadrature.
Here, the 2D CDF at $(x,y)$ is defined by the enclosed mass fraction within the rectangle with diagonal points from $(0,0)$ to $(x,y)$. For $\Delta \phi _{\rm star}$ ($\Delta \phi _{\rm dust}$), we similarly compute the Cram\'{e}r-von Mises statistic $\omega$ using the 2D CDF in the $\phi$ -- radius plane, where $\phi$ represents the azimuthal angle of stellar (or dust) particles. By construction, our toy models are axisymmetric and thus feature uniform $\phi$ coverage at each radius. For a visual representation of the Cram\'{e}r-von Mises statistic see Appendix\ \ref{app:cramer}.

Having introduced the recovery residuals and model mismatch metrics, we now proceed to inspect their interrelation.  Figure\ \ref{fig:residuals} shows the recovery residuals $\mathcal{R}$ as a function of $\Delta {\rm SFH}$, $\Delta {\rm Metal}$, $\Delta {\rm Rad}$, and $\Delta \phi$.  Within each panel, we quote the Spearman rank correlation coefficient $\rho$. For any associated $p$ value smaller than 0.05, indicating the recovery residuals feature a significant correlation with the degree of model mismatch, we format the Spearman rank correlation coefficient $\rho$ in bold font. Among the three redshifts considered, we find that TNG galaxies at z=0 have the overall largest model mismatch but the lowest recovery residuals in a median sense, albeit with a large overlap in distributions. The $z=0$ model mismatch metrics do not correlate strongly with the recovery residuals, perhaps suggesting that other factors than model mismatch (e.g., the lack of rest-frame UV coverage) are more important in driving the recovery residuals. For the other two redshifts ($z=1$ and $z=2$), we find that all four types of model mismatch contribute to recovery residuals but it is hard to determine which type of model mismatch plays a more dominant role.

\begin{figure}
    \centering
    \includegraphics[width=\linewidth]{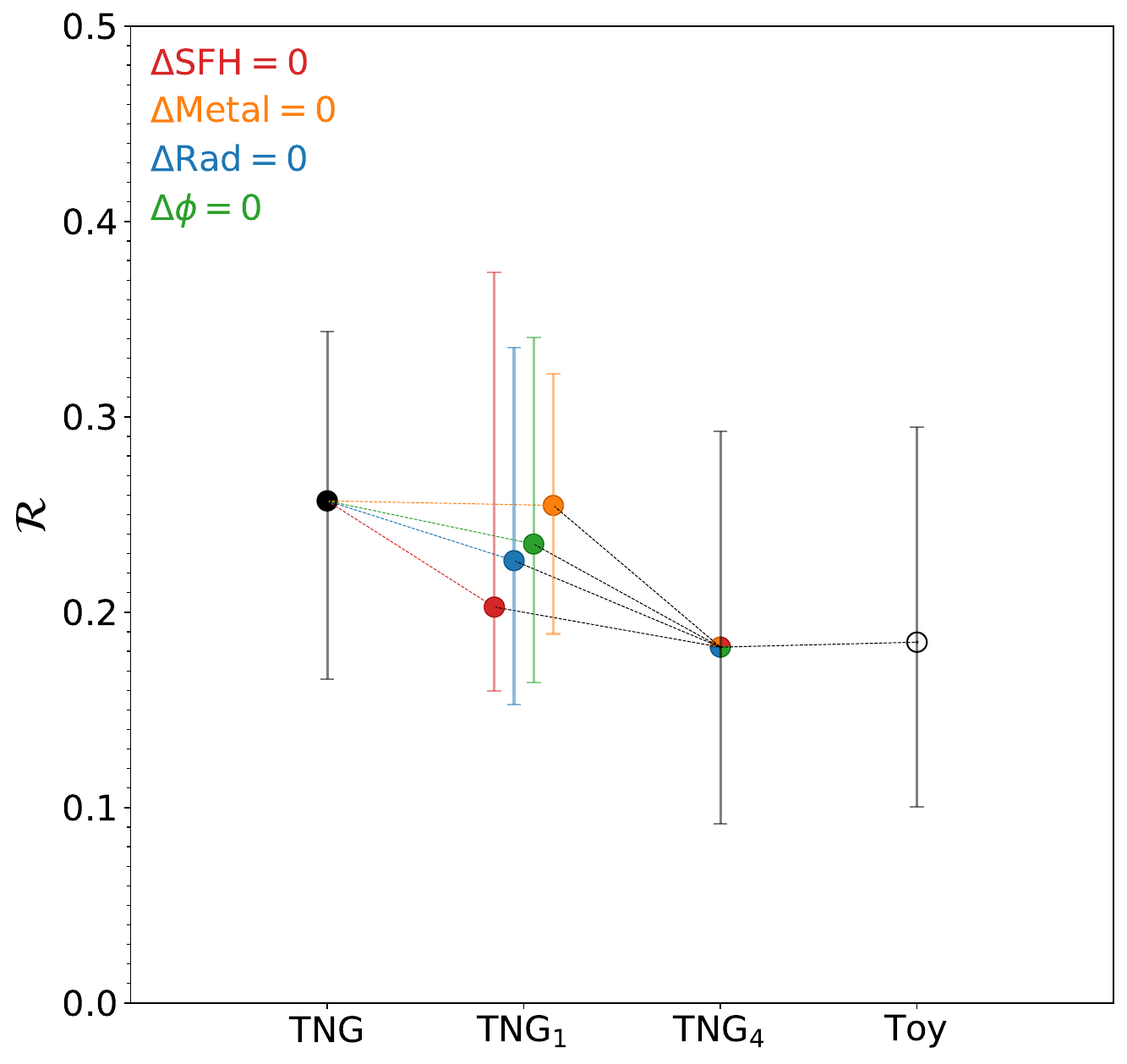}
    \caption{Tests on how recovery residuals $\mathcal{R}$ reduce when artificially simplifying TNG galaxies. Position along the x-axis denotes to what extent the TNG galaxies have been simplified to match the toy model description, from simplifying one of the characteristics shown in the colour legend (labelled as $``\rm{TNG}_1"$), to simplifying all four characteristics ($``\rm{TNG}_4"$), and ultimately a complete toy model galaxy description (``Toy").}
    \label{fig:TNG_simplifications}
\end{figure}

\subsubsection{Tests on artificially simplified TNG galaxies}
\label{sec:simplify}

We now proceed to consider the impact of the same four sources of model mismatch ($\Delta {\rm SFH}$, $\Delta {\rm Metal}$, $\Delta {\rm Rad}$, $\Delta \phi$), but from a different perspective.  For the same set of 16 TNG50 test objects used in Section\ \ref{sec:obssim}, we systematically simplified one characteristic at a time, re-generated mock observations, fitted the resulting SDs with {\tt SE3D} and compared recovery statistics to those obtained for the actual TNG galaxies without modifications. 

For instance, a $\Delta \phi = 0$ simplified TNG galaxy would be generated by randomizing the azimuthal angle of all stellar and dust particles, leaving all other characteristics unchanged.  Similarly, a $\Delta {\rm Metal}$ simplification would be accomplished by resetting all particle metallicities to the mass-weighted stellar metallicity of the galaxy. $\Delta {\rm Rad} = 0$ would involve adjusting the radial coordinates of stellar and dust particles in the $xy$ plane in a rank-preserving manner so that the resulting radial profiles of stars and dust now match the S\'{e}rsic profile of the closest toy model.  Similarly, $\Delta {\rm SFH} = 0$ means the stellar particles' ages are adjusted to match the closest toy model galaxy description while leaving other sources of complexity unchanged.

Figure \ref{fig:TNG_simplifications} illustrates how the median recovery residual $\mathcal{R}$ changes with different simplifications applied to TNG galaxies. For a complete comparison, we also show the recovery statistics for experiments with all four characteristics simplified (labelled as $``\rm{TNG}_4"$) and for the best-approximation toy model.
From this exercise, we find that $\Delta {\rm{SFH}}$ plays the most important role in template mismatch. $\Delta {\rm{Rad}}$ and $\Delta \phi$ also influence the recovery statistics $\mathcal{R}$, but play a more minor role. Eliminating metal mismatch (i.e., imposing $\Delta {\rm{Metal}}=0$) does not have a significant impact on $\mathcal{R}$ for our test objects.  The reason that $\Delta {\rm SFH}$ did not prominently stand out as the main culprit of model mismatch through its correlation with the recovery residuals (Figure\ \ref{fig:residuals}) is likely because the metric quantifies the degree of SFH mismatch regardless of how much this mismatch propagates to the observables.  In other words, a model mismatch in the recent SFH may matter more than one at earlier times. Finally, with simplifications of all four characteristics applied to TNG galaxies (TNG$_4$ in Figure\ \ref{fig:TNG_simplifications}), $\mathcal{R}$ becomes similar to the recovery performance achieved on mock observations generated from a pure toy model galaxy description.  This indicates that the four characteristics considered in this section represent the major sources of model mismatch.

\subsection{Scope for extending {\tt SE3D}}
\label{sec:extensions}

A number of extensions to {\tt SE3D} in its current incarnation could be envisioned with an eye on providing a yet more flexible family of toy models to capture the variety of conditions encountered among real galaxies.  Among them are for instance a volume filling factor for the diffuse dust distribution, or leaving the lifetime of birth clouds as a free parameter.  The use of piecewise constant SFHs\footnote{These are also known as non-parametric SFHs, although somewhat of a misnomer.  As illustrated by \citet{Leja2019}, they too can be sensitive to the precise implementation of priors, especially in applications to broadband photometry; e.g., related to the degree of regularization of the SFH.} has been advocated in SED modelling literature as well, as a means to decouple short- and long-term SFHs and avoid implicit priors on galaxies' specific star formation rate \citep{Leja2019,Carnall2019}.  Radial gradients for such histories could still be implemented via a single parameter, for instance controlling a radial scaling to the time axis, but more complex implementations could of course be formulated as well.

In common to all of the above, an increased dimensionality of the toy model description would require (a) a larger RT training library to properly train the ML emulator (see Figure 5 in R26); (b) an adequate sampler to find the global maximum likelihood solution; and (c) sufficiently constraining data.  It is conceivable that the presently considered SDs, capturing how the integrated light and global structural parameters vary with wavelength, may no longer be sufficient.  Constraints based on the actual multi-wavelength imagery, as opposed to the photometry and global structural parameters, may be more fruitful (see for example the discussion in R26), but a proper joint treatment of mixed resolution (and partially unresolved) observations will remain essential to optimally leverage the available panchromatic information.

\section{Conclusions}
\label{sec:summary}

In this paper, we investigated the ability to recover the intrinsic distribution of dust and stars (+ stellar populations) from panchromatic resolved observations.  We do this by applying our newly developed {\tt SE3D} algorithm, with at its core a ML emulator trained on a library of toy model galaxies that were post-processed using 3D dust radiative transfer.  The inputs to the recovery tests are, for our fiducial settings, $U$-to-870$\mu$m galaxy-integrated photometry and Spectral Structural Distributions capturing the size, S\'{e}rsic index and projected axis ratios of the galaxies measured at multiple wavelengths (see Table\ \ref{tab:filters}).

We generated mock observations (using SKIRT; \citealt{Camps2020}) and carried out the recovery exercise for two types of test objects: parametrized toy models and massive SFGs extracted from the TNG50 cosmological simulation.  The toy models by design span a wide range in stellar populations and star-dust geometries.  The TNG50 SFGs feature an inherently more complex make-up.  We compute their age gradient, dust-to-stellar size ratio and thickness of the stellar and dust distribution, and present the evolution of these properties from $z=2$ to the present day among the mass-selected ($\log(M_{\rm star}~[M_\odot]) > 10$) SFG sample.  Massive SFGs in TNG feature predominantly negative age gradients, particularly so at later times.  Nominal procedures to assign dust based on the mass and metallicity of cold/dense gas particles result in dust distributions that are more radially extended than the stars, by factors $\sim 1.6$ (at $z=2$) to $\sim 2$ (at $z=0$), with significant object-to-object variation.  We recover a pattern of disk settling, in terms of a decreasing geometric thickness of stellar and dust distributions since cosmic noon, with at any epoch the dust distribution being thinner (i.e., $C_{\rm dust}/ R_{\rm dust} < C_{\rm star}/R_{\rm star}$).

In observable space, we find the mock-observed TNG50 SFGs to cover a relatively modest dynamic range in the $UVJ$ diagram, lacking an extension to the most dust-reddened SFGs in observed samples, in line with previous findings in the literature.  Also in IRX-$\beta$ there are signs of a relatively inefficient dust reddening despite an appreciable range of attenuation levels and IR-to-UV luminosity ratios (although none of them extreme).  We identified a lack of redshift evolution in $M_{\rm dust}/M_{\rm star}$ and differences in star-dust geometries as potential culprits for the different distributions in $UVJ$ and IRX-$\beta$.  In comparison, our library of toy model galaxies covers a much wider range across $UVJ$ and IRX-$\beta$, encompassing the regions occupied by observed galaxy samples.  For this reason, and with the intention to test {\tt SE3D} in the absence and presence of template mismatch, we applied our recovery analysis on mock observations of both toy model and simulated nature.

Despite finite wavelength sampling and the introduction of measurement errors, the recovered properties for toy models correspond well to the intrinsic truth.  Deviations in stellar and dust mass as well as SFR are on the order of $\sim 0.1$ dex, with negligible systematic offsets.  A similar recovery quality is achieved for the half stellar mass, half dust mass and half SFR radii.  The mass-weighted age and its spatial gradient exhibit somewhat more scatter, but again without systematic offsets.  Stellar metallicities and birth cloud covering fractions are more poorly constrained, although in the absence of model mismatch the correlation between recovered and intrinsic properties is still of high statistical significance ($\rho \sim 0.4 - 0.8;\ p < 10^{-5}$).

Tests with more limited input data (i.e., restricting wavelength coverage, or wavelength coverage with resolved information) were conducted, showing encouraging performance for the typical data availability in legacy deep fields.  Interestingly, we demonstrate that a rudimentary assessment of galaxy scale (with recovery of stellar, dust and SFR sizes to $\sim 0.2 - 0.3$ dex precision) is feasible from the SED information alone, provided the object studied has a similar physical make-up as the toy model description we employ in our modelling.  This owes to the fact that panchromatic SEDs carry information on both bulk dust content (in the Rayleigh-Jeans tail) and dust columns (via reddening at short wavelengths), and because additionally more compact configurations translate to higher dust temperatures.

For mock-observed SFGs in TNG50, the match between recovered and real properties remains equally good as for toy models when considering the mass and size of the stellar and dust components, or when considering the total SFR and half-SFR radius, with recovery accuracies around $\sim 0.1$ dex.  The recovery of mass-weighted ages or age gradients is comparatively poorer, albeit still sufficient to trace evolutionary trends among the SFG population across cosmic time.  Where scatter in recovered properties is enhanced, we attribute this to model mismatch (i.e., complex TNG galaxies being more poorly represented by our simplistic toy model description).  An investigation of which source of model mismatch matters most suggests that SFH plays a major role while radial profile shape and azimuthal structure represent more minor contributions to template mismatch. Replacing the stellar metallicity distribution with a delta function at the mass-weighted metallicity does not impact the goodness of recovery.

Panchromatic observations are increasingly available for distant galaxy samples, with sufficient resolution in multiple bands enabling the measurement of global structural parameters for different tracers (young/old stars, PAH and cold dust emission).  {\tt SE3D} offers a framework to jointly analyse such observational diagnostics.  It propagates the observables (wavelength-dependent fluxes, sizes, S\'{e}rsic indices and projected axis ratios) from an input 3D configuration of dust and stars in a self-consistent manner, accounting for the observer's viewing angle.  The impact of dust on observables is thus not imposed in the form of an effective attenuation law, rather the effective attenuation and reddening emerges from the input physical conditions.  The same goes for dust temperature, which is not treated as a free parameter.  Likewise, a balance between absorbed and re-emitted energy, while preserved when integrating over $4\pi$ steradians, is not imposed for an individual observer's perspective.

We demonstrate an encouraging performance for a number of key physical parameters, point out the limitations of broadband diagnostics to constrain others, and caution for the effects of model mismatch.  We also highlight how mock observational diagnostics for cosmologically simulated galaxies do not necessarily span the full range exhibited by real galaxies, a shortcoming which may inhibit certain forms of direct, simulation-based inference of observed galaxies' physical properties.

\vspace{-0.1cm}
\section*{Acknowledgements}

We thank the authors of SKIRT, Maarten Baes and Peter Camps, for making their radiative transfer code publicly available.  We also thank James Trayford, Andrea Gebek, Nick Andreadis, Shiyin Shen and XianZhong Zheng for valuable discussions on this work.  The authors gratefully acknowledge support from the Royal Society International Exchanges scheme (IES\textbackslash R2\textbackslash 242195).  SW acknowledges support from China's National Foreign Expert programme (H20240871).  The authors acknowledge the Tsinghua Astrophysics High-Performance Computing platform at Tsinghua University for providing computational and data storage resources that have contributed to the research results reported within this paper.

\vspace{-0.1cm}
\section*{Data Availability}

The TNG50 simulation data used in this work can be accessed at \url{https://www.tng-project.org/} and is further documented in \citet{Nelson2019b}.  The observational samples shown for reference are extracted from the DAWN JWST Archive (\url{https://dawn-cph.github.io/dja/blog/2024/08/16/morphological-data/}), with ancillary information taken from legacy {\it Spitzer} and {\it Herschel} surveys in the same fields (see Section\ \ref{sec:obsdiagrams}).



\bibliographystyle{mnras}
\bibliography{bibliography} 




\appendix
\section{Alternative approaches to introduce dust in TNG}
\label{app:dustassignment}

\begin{figure*}
    \centering
    \includegraphics[width=0.6\linewidth]{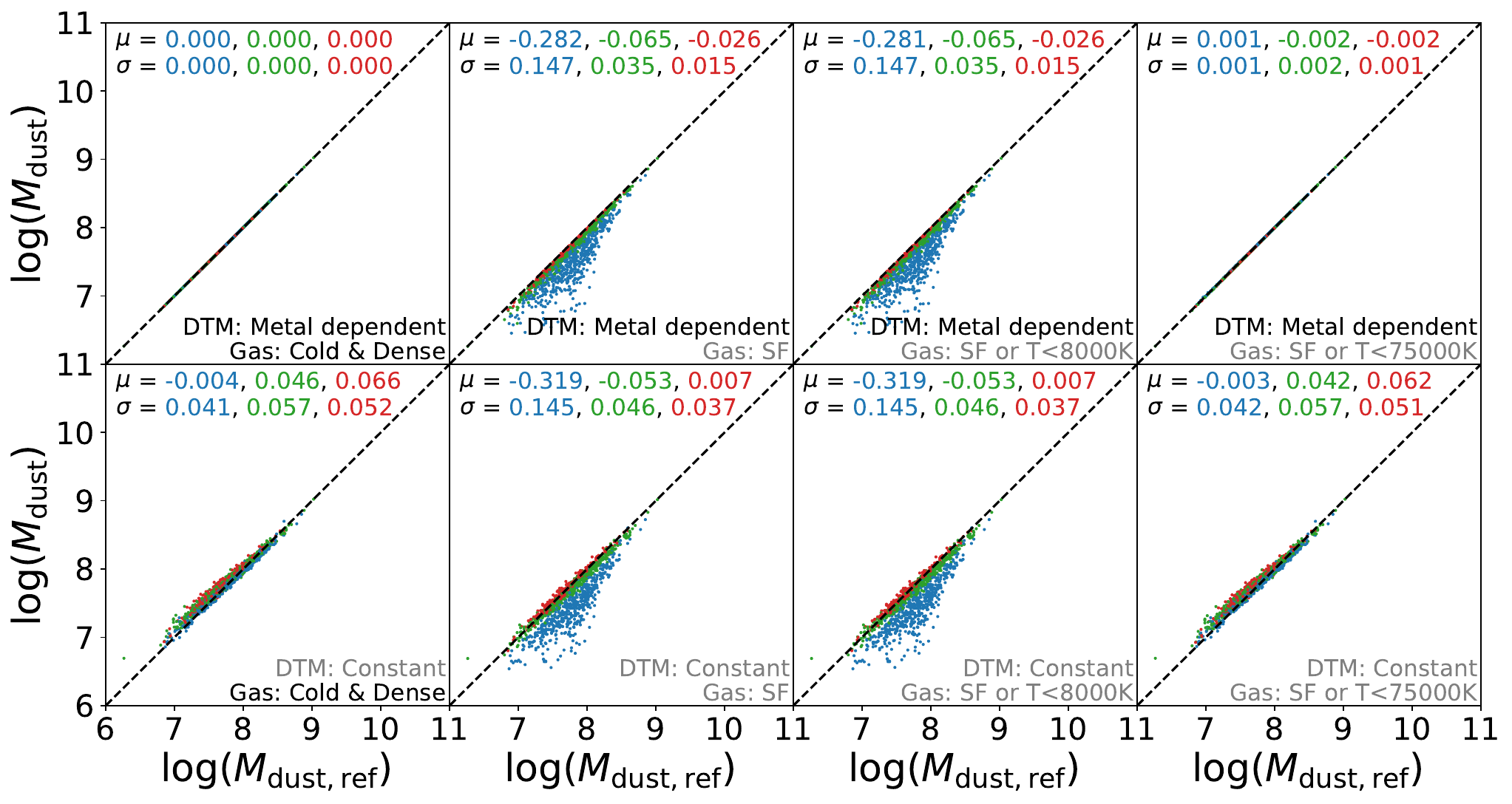}
    \includegraphics[width=0.6\linewidth]{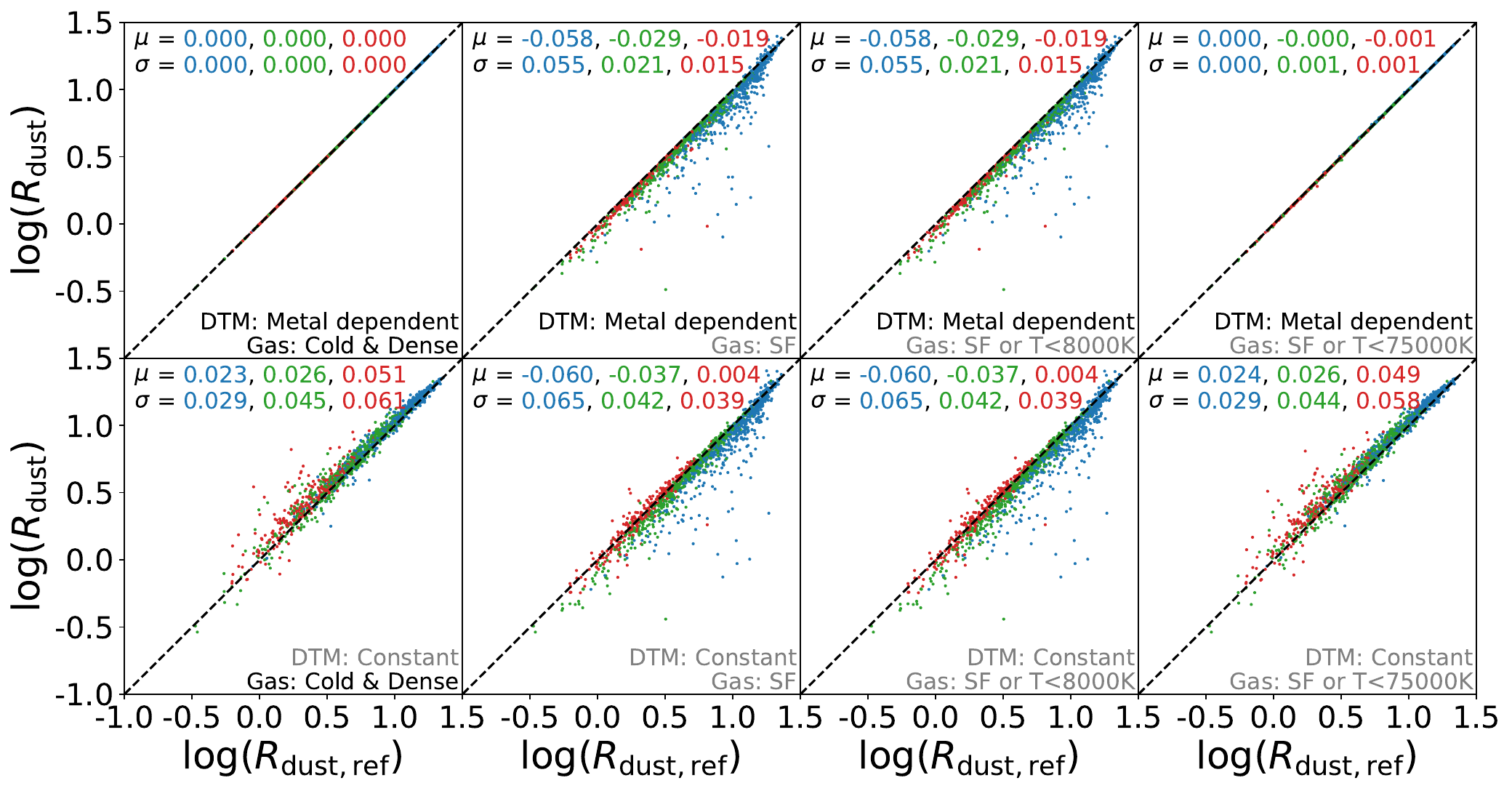}
    \includegraphics[width=0.6\linewidth]{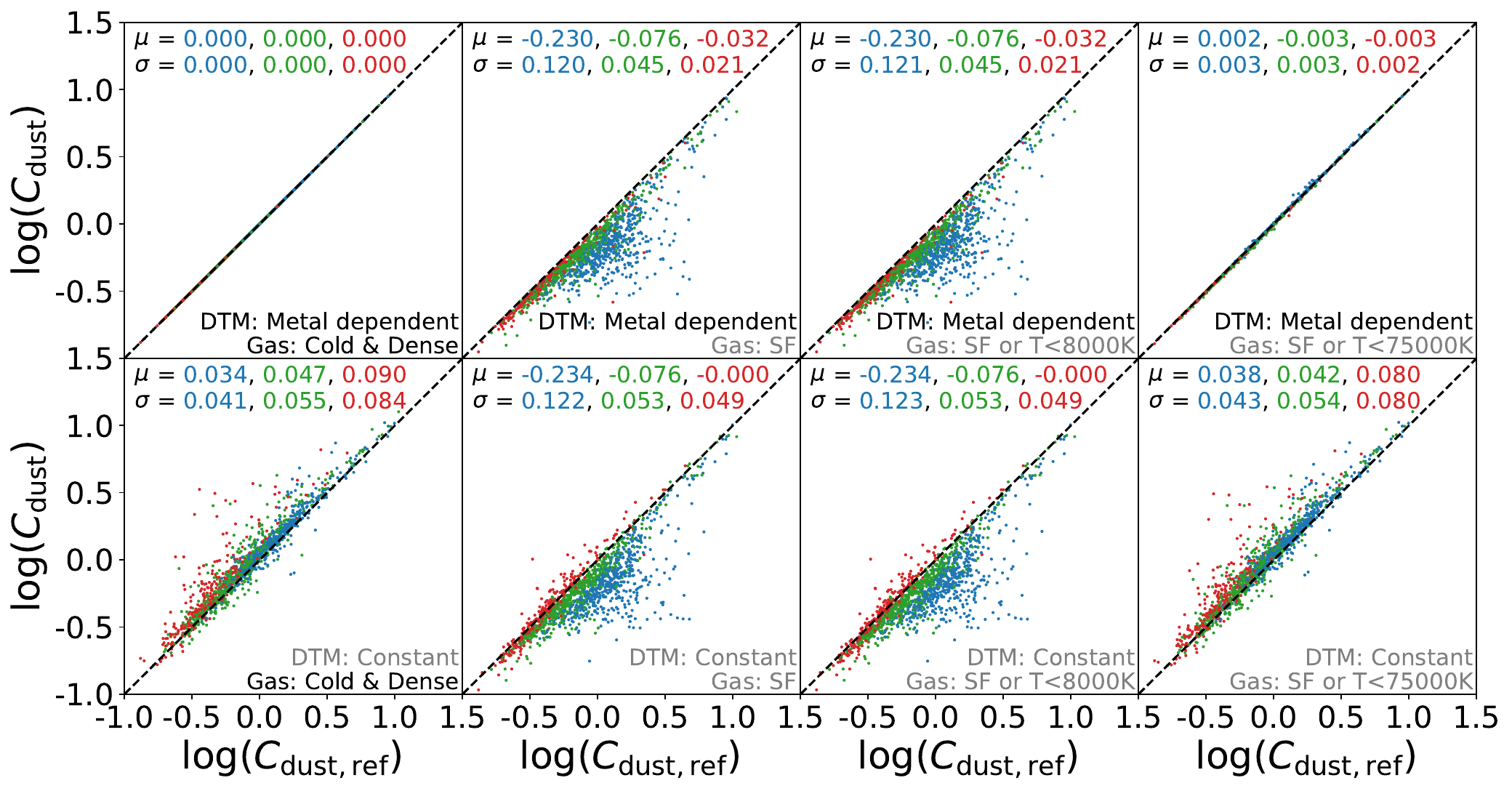}
    \includegraphics[width=0.6\linewidth]{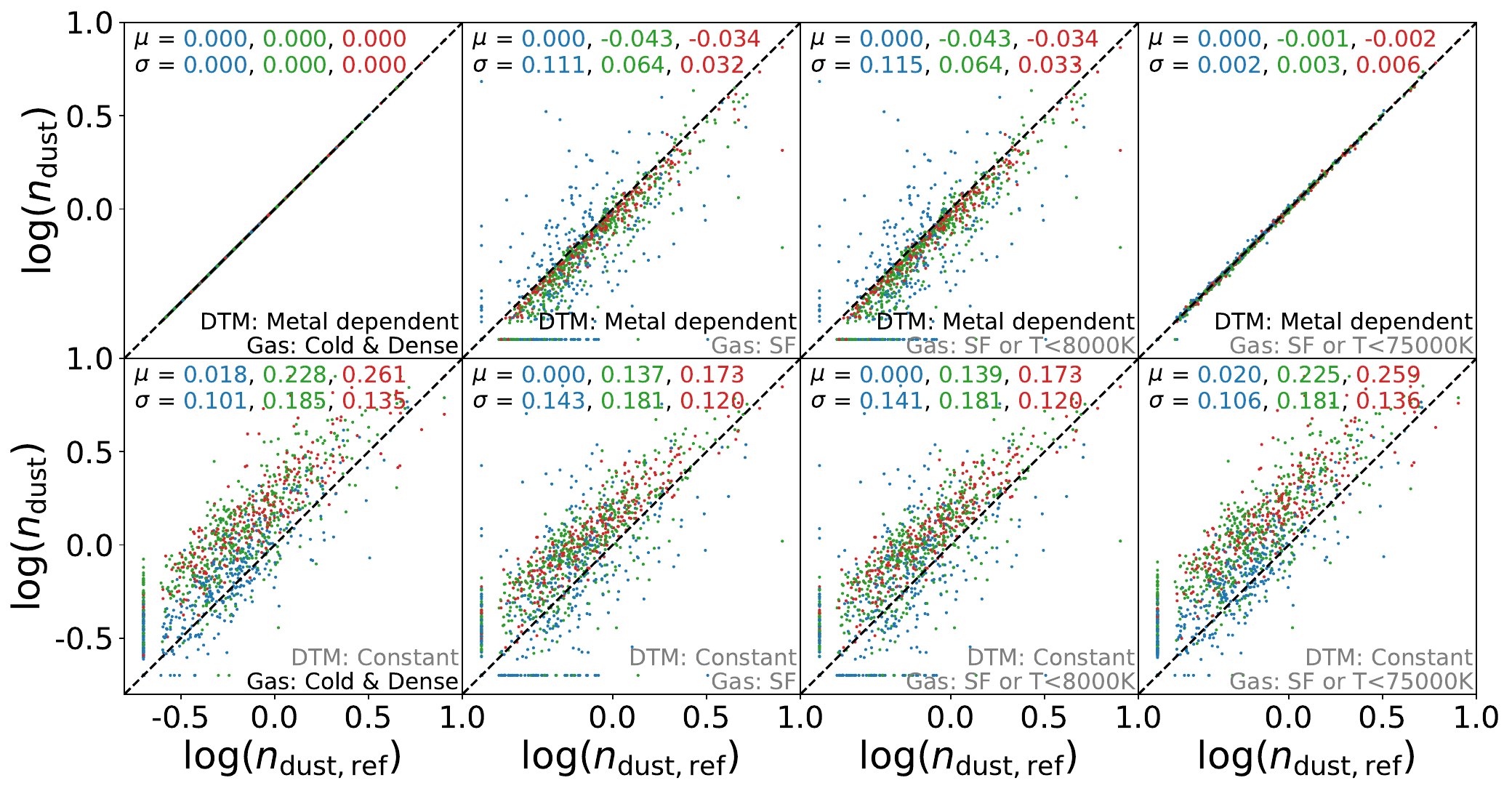} 
    \caption{Comparison of dust properties for different methods of assigning dust to gas particles in TNG50.  Each property on the x-axis corresponds to that obtained using our default (i.e., reference) method.  We evaluate the impact on total dust mass ($M_{\rm dust}$), radius ($R_{\rm dust}$) and height ($C_{\rm dust}$) of the dust distribution, as well as the S\'{e}rsic index ($n_{\rm dust}$) best describing its radial profile shape.  The median offset ($\mu$) and scatter ($\sigma$) are quoted in each panel for $z = 0$ (blue), $z=1$ (green) and $z=2$ (red).  We consider both variations in the assumed dust-to-metal ratio (DTM) and to which gas particles dust is assigned.}
    \label{fig:TNG_assign_gas_dust}
\end{figure*}

Cosmological simulations such as TNG do not explicitly incorporate dust physics nor do they resolve the multi-phase ISM.  Instead, they adopt an effective equation of state and physical model for cooling in which gas is not allowed to reach temperatures significantly below $10^4$ K \citep[see, e.g.,][]{Pillepich2018a}.  For this reason, any RT post-processing requires a decision on which gas particles to assign dust to, and what amount.  Various approaches have been adopted in the literature.  Our default approach is outlined in Section\ \ref{sec:assigningdust}.  It follows \citet{Torrey2012} and \citet{Gebek2025} to identify the relevant gas particles on the basis of a selection in the temperature -- density plane, and adopts the empirical dust-to-metal (DTM) calibration by \citet{DeVis2019}.  For the selection of dust-hosting gas particles, alternatives in the literature include those gas particles that are star-forming \citep[e.g.,][]{Rodriguez-Gomez2019}, optionally supplemented by any gas particles below a temperature threshold (e.g., $T < 8000 {\rm K}$ by \citealt{Camps2016}, \citealt{Trayford2017} and \citet{Vogelsberger2020}; $T < 75,000 {\rm K}$ by \citealt{Schulz2020} and \citealt{Popping2022}).  Several of these works adopted a constant dust-to-metal ratio of ${\rm DTM} = 0.3$ \citep{Camps2016,Trayford2017,Schulz2020}, although also higher values of ${\rm DTM} = 0.4$ have been used \citep{Roebuck2019}. \citet{Popping2022} instead worked with the metallicity-dependent prescription by \citet{Remy-Ruyer2014}, and \citet{Donnari2019} and \citet{Vogelsberger2020} adopted a redshift-dependent DTM.  

In Figure\ \ref{fig:TNG_assign_gas_dust} we explore the impact of such choices on the bulk amount of dust and on its global structural properties.  We contrast our fiducial model to ones that vary in the selection of gas particles to which dust is assigned (columns) and ones with a constant ${\rm DTM} = 0.3$ (bottom row for each dust property).  The impact of different gas particle selection methods is small at $z = 1$ and $z = 2$, but more substantial at $z=0$, with a reduction in $M_{\rm dust}$ by $\sim 0.3$ dex for some alternative methods.  The distinction between a constant DTM and our fiducial approach can be understood from the fact that our empirically calibrated, metallicity-dependent DTM scaling assigns relatively less dust to metal-poor gas than the constant DTM case.  Adopting a constant DTM would therefore increase the total dust masses by 0.07 dex in the median (more pronounced among low-$M_{\rm dust}$ objects as they tend to be more metal poor).  Since the change is mostly to metal-poor gas which resides in the galaxy outskirts, it further has the consequence of increasing the radial and height distribution of the dust (by 0.06 and 0.09 dex, respectively), as well as its S\'{e}rsic index, which is sensitive to the profile wings (by 0.25 dex).

\begin{figure*}
    \centering
    \includegraphics[width=\linewidth]{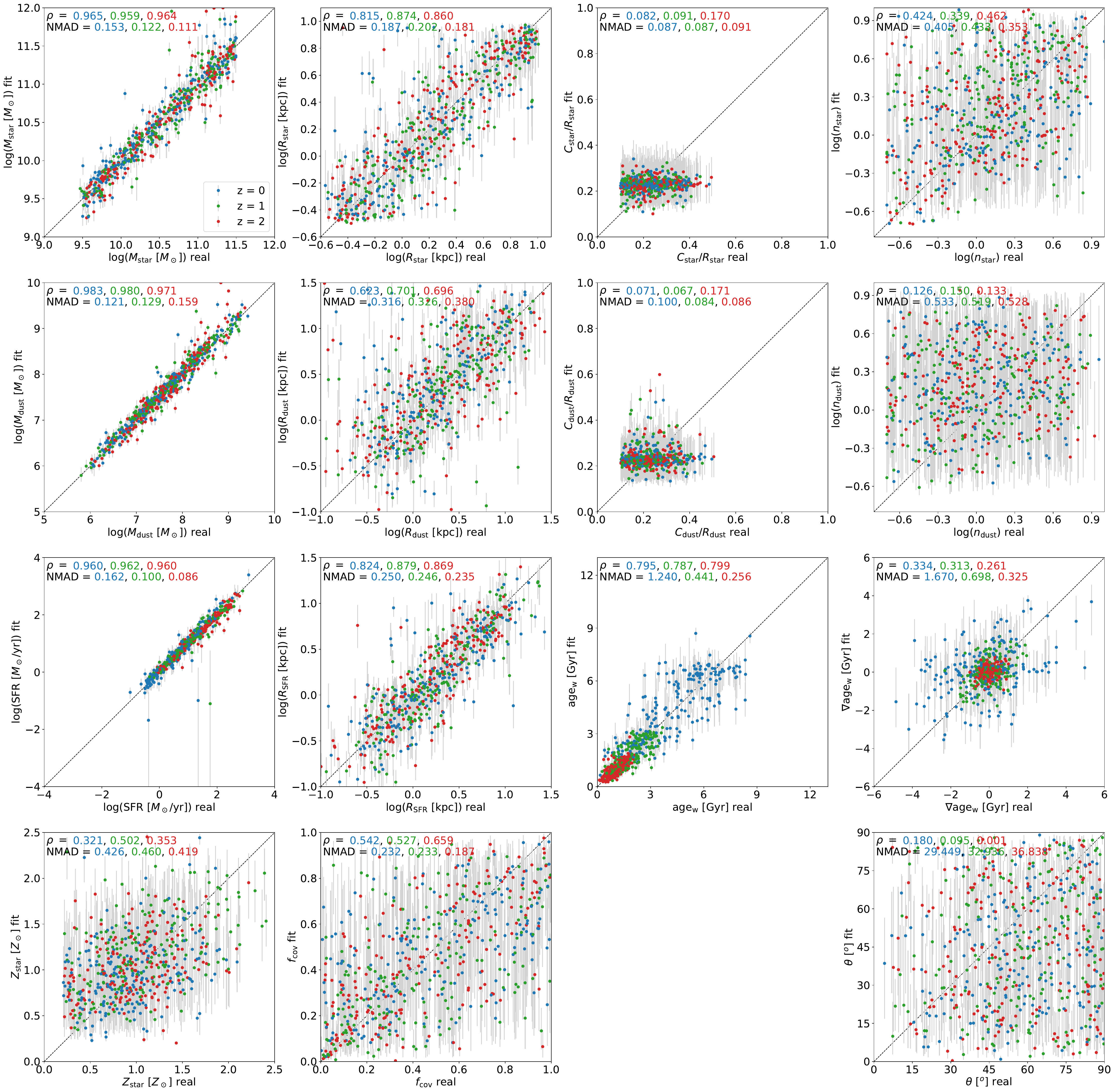}
    \caption{Recovered parameters from {\tt SE3D} fitting contrasted with true intrinsic toy model parameters for the filter band combination `nores' (i.e., lacking any spatial information) and 0.05 dex uncertainty. We fit a sample of 600 toy model galaxies equally divided between three redshifts: $z=0$ (blue), $z=1$ (green), and $z=2$ (red).}
    \label{fig:SE3Dfit_nores}
\end{figure*}

We verified on a dozen test objects that systematic changes to the $UVJ$ and IRX-$\beta$ observables investigated in Section\ \ref{sec:obsdiagrams} propagating from such alternative approaches remain limited to $\sim -0.031$ in $U-V$, $\sim -0.065$ in $V-J$, $\sim -0.024$ in $\beta$ and $\sim -0.052$ in IRX.

\section{Recovery on mock-observed toy models without resolved information}
\label{app:nores}

Figure\ \ref{fig:SE3Dfit_nores} presents results from {\tt SE3D} fitting to panchromatic SEDs of toy model galaxies, without the use of resolved information (i.e., the `nores' setting detailed in Table\ \ref{tab:filters}).  Noteworthy is that size estimates with scatter on the order of $\sim 0.2 - 0.3$ dex are retrieved, despite the lack of resolved diagnostics among the input observables.  Far-IR constraints on dust mass and temperature along with the imprint of dust reddening and attenuation at shorter wavelengths enable an assessment of how extended or compact a galaxy is.

\begin{figure*}
    \centering
    \includegraphics[width=\linewidth]{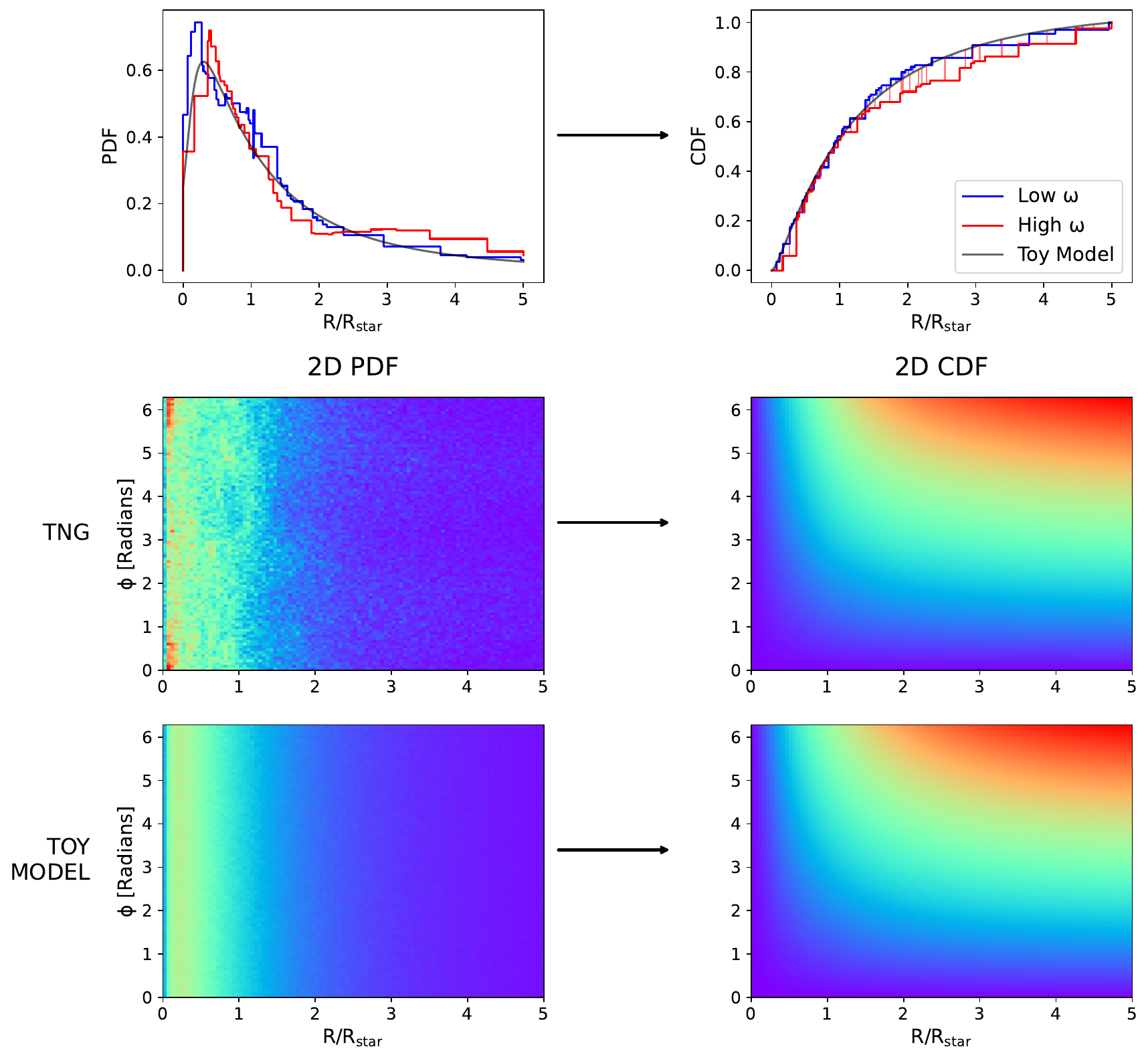}
    \caption{Visual representation of the Cram\'{e}r-von Mises statistic $\omega$ for radial model mismatch ({\it top row}) and azimuthal model mismatch ({\it bottom two rows}). The top row displays a 1D PDF and corresponding CDF for a low and high $\omega$ TNG galaxy compared to our toy model description (which has a stellar S\'{e}rsic index of $n = 2.4$).  The second and third row display the 2D PDF and CDF of the stellar distribution in azimuthal angle versus radius space, for the low $\omega$ TNG galaxy compared to our toy model.  The Cram\'{e}r-von Mises statistic sums over the squared residuals between TNG and toy model CDFs (Eq.\ \ref{eq:CvM}).}
    \label{fig:Cramer_von_Mises}
\end{figure*}

\section{Cram\'{e}r-von Mises}
\label{app:cramer}
Figure \ref{fig:Cramer_von_Mises} displays a visual representation of the ingredients to compute the Cram\'{e}r-von Mises statistic $\omega$, as described by Equation\ \ref{eq:CvM}.  The top row illustrates the 1D case of radial model mismatch for the stellar distributions of two example TNG galaxies (in blue and red) with respect to the best-fitting toy model description, which in both cases has a S\'{e}rsic index $n = 2.4$.  The larger CDF residuals for the TNG galaxy displayed in red translate via Equation\ \ref{eq:CvM} to a higher value of $\omega$.  For illustrative purposes, the probability distribution function (PDF) and corresponding cumulative distribution function (CDF) for the TNG galaxies were heavily downsampled.  The actual $\omega$ values used in our analysis are computed on the full CDFs, which contain as many steps as the TNG galaxy contains stellar particles.

The bottom two rows illustrate the azimuthal model mismatch discussed in Section \ref{sec:accuracy_vs_mismatch}, for the TNG galaxy whose radial model mismatch is displayed in blue in the top row. By construction, the toy model description lacks azimuthal variations.  The summation (in quadrature) of CDF residuals described by Equation\ \ref{eq:CvM} is now taken over all bins of the 2D CDF diagrams depicted in the right-hand panels.


\bsp	
\label{lastpage}
\end{document}